\documentclass[useAMS,usenatbib]{mn2e}
\usepackage[dvipdfmx]{color,graphicx}
\usepackage{multirow}
\def\aj{AJ}             	
\def\apj{ApJ}           	
\def\apjl{ApJ}          	
\def\apjs{ApJS}         	
\def\aap{A\&A}          	
\def\mnras{MNRAS}       	
\def\pasp{PASP}         	
\def\nat{Nature}        	

\def\Msun{~{\rm M_{\odot}}}

\def\gsim{\hspace{0.3em}\raisebox{0.4ex}{$>$}\hspace{-0.75em}\raisebox{-.7ex}{$\sim$}\hspace{0.3em}}
\def\lsim{\hspace{0.3em}\raisebox{0.4ex}{$<$}\hspace{-0.75em}\raisebox{-.7ex}{$\sim$}\hspace{0.3em}}

\title[Non-linear instability for clump formation]{Non-linear violent disc instability with high Toomre's $Q$ in high-redshift clumpy disc galaxies}
\author[S. Inoue et al.]
{\parbox[t]{\textwidth} 
{Shigeki Inoue$^{1}$\thanks{E-mail: shigeki.inoue@mail.huji.ac.il}, Avishai Dekel$^{1}$, Nir Mandelker$^{1}$, Daniel Ceverino$^{2,3}$, Fr\'ed\'eric Bournaud$^{4}$ and Joel Primack$^{5}$}
\\ \\
$^{1}$Racah Institute of Physics, The Hebrew University, Jerusalem, 91904, Israel\\
$^{2}$Centro de Astrobiolog{\'i}a (CSIC-INTA), Ctra de Torrej{\'o}n a Ajalvir, km 4, E-28850 Torrej{\'o}n de Ardoz, Madrid, Spain\\
$^{3}$Astro-UAM, Universidad Autonoma de Madrid, Unidad Asociada CSIC, E-28049 Madrid, Spain\\
$^{4}$Laboratoire AIM-Paris-Saclay, CEA/DSM-CNRS-Universi\'e Paris Diderot, Irfu/Service d'Astrophysique, CEA Saclay,\\ Orme des Merisiers, F-91191 Gif sur Yvette, France\\
$^{5}$Department of Physics, University of California, Santa Cruz, CA 95064, USA
}

\date{2014 May 25}

\begin{document}

\pagerange{\pageref{firstpage}--\pageref{lastpage}} \pubyear{2014}

\maketitle

\label{firstpage}

\begin{abstract}

We utilize zoom-in cosmological simulations to study the nature of violent disc instability (VDI) in clumpy galaxies at high redshift, $z=1$--$5$.  Our simulated galaxies are not in the ideal state assumed in Toomre instability, of linear fluctuations in an isolated, uniform, rotating disk. There, instability is characterised by a $Q$ parameter below unity, and lower when the disk is thick. Instead, the high-redshift discs are highly perturbed. Over long periods they consist of non-linear perturbations, compact massive clumps and extended structures, with new clumps forming in inter-clump regions. This is while the galaxy is subject to frequent external perturbances. We compute the local, two-component $Q$ parameter for gas and stars, smoothed on a $\sim1~{\rm kpc}$ scale to capture clumps of $10^{8-9}~{\rm M}_\odot$. The $Q<1$ regions are confined to collapsed clumps due to the high surface density there, while the inter-clump regions show $Q$ significantly higher than unity. Tracing the clumps back to their relatively smooth Lagrangian patches, we find that $Q$ prior to clump formation typically ranges from unity to a few. This is unlike the expectations from standard Toomre instability. We discuss possible mechanisms for high-$Q$ clump formation, e.g. rapid turbulence decay leading to small clumps that grow by mergers, non-axisymmetric instability, or clump formation induced by non-linear perturbations in the disk. Alternatively, the high-$Q$ non-linear VDI may be stimulated by the external perturbations such as mergers and counter-rotating streams. The high $Q$ may represent excessive compressive modes of turbulence, possibly induced by tidal interactions.

\end{abstract}


\section{Introduction}
\label{Intro}
Observations for the Hubble Deep Field have revealed clumpy morphology of galaxies in the high-redshift Universe \citep[][]{bae:96}. Some of the clumpy galaxies were observed to have disc structures characterized by significant rotations \citep[e.g.][]{g:06,g:08} and were interpreted as early disc galaxies \citep[e.g.][]{n:98,n:99,ab:01,eeh:04} although some fraction of them may be ongoing mergers \citep[e.g.,][]{w:06,fgb:09,p:10}. The face-on and edge-on views of the clumpy discs were referred to as clump clusters and chain galaxies \citep{eeh:04}. The observed discs host giant clumps with masses $M_{\rm cl}\lsim 10^9{\rm M_\odot}$. Recent observations showed that nearly half the galaxies of stellar masses from $10^9$ to $10^{11.5}~{\rm M_\odot}$ have clumpy morphology at redshifts of $z\simeq2$--$3$ \citep{tkt:13II,mkt:14,gfb:14}. Clumpy disc galaxies have also been observed in the present Universe, but their abundance is much lower \citep[e.g.][]{ees:13,bgf:14,gpm:15}.

In parallel, numerical simulations have been very instrumental in predicting the formation processes of galaxies at high redshifts and in understanding their nature. Before the discovery of clumpy galaxies in the Hubble Deep Field, using isolated galaxy models, \citet{SN:93} predicted that, if disc galaxies are highly gas-rich in their formative stages, they would form giant clumps in their discs by Toomre instability \citep{t:64}. Subsequent theoretical studies elucidated various aspects of the clumpy galaxy formation scenario. The massive clumps migrate into the galactic centre and can form a bulge \citep[e.g.][]{n:98,n:99,g:06,g:08,ebe:08,dsc:09,is:12,pvt:13,cdt:14}. The clumps can affect the density distribution of the dark matter \citep{ebe:08,is:11} and the kinematic state of stellar halo objects such as halo stars and globular clusters \citep{i:13}. Additionally, the clumpy states of galactic discs could be a possible formation scenario for thick discs (\citealt[][]{n:96,bem:09}, although see \citealt{is:14}), the exponential density profiles of stellar discs \citep{eev:05,bee:07,es:13} and the metal-rich globular clusters associated with bulges and discs \citep{sgf:10}.


\citet{lmk:05a,lmk:05b,lmk:06} have performed numerical simulations using isolated disc models in which the formation and growth of star clusters are represented by sink particles, and they showed that the formation of star clusters is indeed controlled by Toomre instability via the parameter of $Q$. Observations for the large Magellanic cloud by \citet{ygc:07} also found that the abundance of young stellar objects increases exponentially as $Q$ decreases. Therefore, if giant clumps in high-redshift galaxies are scaled-up versions of star clusters in the local galaxies, we could expect that high-redshift clump formation is also correlated with the value of $Q$. 

\citet{gnj:11,gfl:14} have observed clumpy disc galaxies at redshifts $z\sim2$ using SINFONI/VLT integral field spectroscopy and estimated the two-dimensional distribution of $Q$ from the gas component of the non-linearly perturbed discs. They found $Q<1$ in the extended disc regions, as opposed to the inner regions that are dominated by bulges. This seems to be consistent with standard linear Toomre instability in an outer ring and morphological quenching inside \citep{mbt:09}. In addition, \citet{fgb:14} and \citet{ogb:15} observed clumpy disc galaxies in the nearby Universe and found that their galaxies have $Q\lsim1$ although they used a global model to estimate $Q$. In the local Universe, the solar neighbourhood in our Galaxy and nearby spiral galaxies that show no clumpy structures have been generally observed to be marginally stable in terms of Toomre instability with $Q=1$--$3$ \citep[e.g.][]{r:01,lwb:08,flw:14,wab:14}. The values of $Q$ in disc galaxies have been discussed to increase with time by theoretical studies \citep{dsc:09,cdg:12}, possibly indicating evolution from unstable to stable states of the galaxies if the high-redshift clumpy discs are progenitors of the local spiral galaxies.

This paper presents results of Toomre analysis of clumpy disc galaxies in the Vela data set of zoom-in cosmological simulations using an adaptive mesh refinement (AMR) code \citep[e.g.][]{ckk:14,z:15}. Our computations of $Q$ are based on the two-component model of \citet{rw:11} and use locally measured quantities smoothed on a scale of $\simeq1~{\rm kpc}$. Details of our simulations and the sample of galaxies used are described in \S\ref{Cos-simulation}. Our method to compute $Q$ is explained in \S\ref{Analysis}, and the results are presented in \S\ref{Results}. We discuss our results in \S\ref{Discussion} and summarize our conclusions in \S\ref{Conclusions}.

\section{Simulations}
\label{Cos-simulation}
The Vela simulations are performed with the {\tt ART} code \citep{kkk:97,k:03,ck:09}. The cosmological parameters are set to be $\Omega_m=0.27$, $\Omega_\lambda=0.73$, $\Omega_b=0.045$, $h=0.7$ and $\sigma_8=0.82$ \citep{komatsu:09}. Each galaxy is resolved with a zoom-in method reaching an AMR maximum resolution of $17.5-35~{\rm pc}$ in re-runs following a low-resolution $N$-body simulation in a large comoving cosmological box.

The simulations include gas cooling by atomic and molecular hydrogen, helium and metals, and heating by the ultra-violet background with partial self-shielding \citep{hm:96}. The cooling and heating rates are based on the {\tt CLOUDY} code \citep{fkv:98}, assuming a slab of thickness of $1~{\rm kpc}$. Our code implements stellar feedback by stellar winds and type-Ia and -II supernovae (SNe) as local injection of thermal energy \citep{ck:09}. The code also incorporates radiative feedback from massive young stars as a non-thermal pressure term \citep{ckk:14}. The supernovae eject mass and metals into the gas cells that contain the stellar particles responsible for the supernovae. Artificial fragmentation of gas is prevented by introducing a pressure floor, which ensures that the Jeans scale is resolved by at least seven cells \citep[see][]{cdb:10}. 

Stellar particles are created in gas cells according to the criteria for hydrogen number density of $n>1~{\rm cm^3}$ and temperature of $T<10^4~{\rm K}$. We use a stochastic model of star formation that yields a star-formation efficiency per free-fall time of nearly two per cent. The initial mass function is assumed to follow the form of \citet{c:05} for each stellar particle. Our cosmological simulations incorporate a thermal stellar feedback model, in which the combined energy from stellar winds and SNe is released as a constant heating rate for $40~{\rm Myr}$ after star formation. The gas cooling is not switched off during SN explosion. The code also implements the effect of runaway stars by assigning a velocity kick of $\sim10~{\rm km~s^{-1}}$ to thirty per cent of the newly formed stellar particles. However, the code does not incorporate the effects of active galactic nuclei, cosmic rays or magnetic fields. Given that we do not resolve the early adiabatic phases of the SNe, that our feedback may be on the low side, and that the feedback of active galactic nuclei is not included, our simulations may underestimate the overall feedback effects \citep{ckk:14,mgm:14}. As a result, stars form somewhat early, leaving the galaxies at $z\sim2$ with gas fractions, star formation rates (SFRs) that are lower than the observed values by up to a factor of $\sim2$. We assume that these inaccuracies do not alter the qualitative behaviors of the simulated galaxies.

The maximum resolution in the zoom-in hydro simulations, which is valid in particular in the cold discs and dense clumps, enables the simulations to resolve gas densities of $\sim10^3~{\rm cm^{-3}}$ with temperatures of $\sim300~{\rm K}$. Each dark-matter particle has a mass of $8.3\times10^4~\Msun$, and each stellar particle has at least a mass of $10^3~\Msun$. The AMR scheme splits a cell into 8 sub-cells when the parent cell contains a mass greater than $2.6\times10^5~\Msun$ in stellar and dark matter particles or $1.5\times10^6~\Msun$ in gas.

\begin{table*}
  \begin{center}
    \begin{minipage}{170mm}
      \caption{Properties of our sample galaxies from the cosmological simulations. The values of $M_{\rm Star}$, $M_{\rm Gas}$, $f_{\rm gas}$, $B/T$ and SFR are measured within $0.15R_{\rm vir}$. The fraction of $f_{\rm gas}$ refers to the mass ratio of gas to total baryon, but $M_{\rm Gas}$ indicates mass of gas plus the stars younger than $100~{\rm Myr}$, while $M_{\rm Star}$ refers to the mass of the other stars. The SFR is averaged within the last $100~{\rm Myr}$. The ratio of $B/T$ is the bulge-to-total stellar mass ratio, where bulge stars obey the criterion of $j_z/j_{\rm max}<0.7$ (see \S\ref{bulge}). The value of $f_{\rm Gas}^{\rm disc}$ is the mass fraction of Gas to total baryons excluding the bulge stars within $|z|<3~{\rm kpc}$. The rightmost column indicates the figures in which the results of the snapshots are shown.}
      \label{CosList}
      $$ 
      \begin{tabular}{ccccccccccc}
        \hline
        \noalign{\smallskip}
        \multirow{2}{*}{run} & \multirow{2}{*}{redshift} & $M_{\rm vir}$ & $R_{\rm vir}$ & $M_{\rm Star}$ & $M_{\rm Gas}$ & \multirow{2}{*}{$f_{\rm gas}$} & \multirow{2}{*}{$f_{\rm Gas}^{\rm disc}$} & \multirow{2}{*}{$B/T$} & SFR & \multirow{2}{*}{figures} \\
        & & ${\rm [M_\odot]}$ & ${\rm [kpc]}$ &  ${\rm [M_\odot]}$ & ${\rm [M_\odot]}$ & & & & ${\rm [M_\odot~yr^{-1}]}$  &\\
        \hline
        V07 & $2.13$ & $8.8\times10^{11}$ & $100$ & $5.6\times10^{10}$ & $1.6\times10^{10}$ & $0.18$ & $0.37$ & $0.37$ & $27.5$ & Fig. \ref{DensityMaps_V07_z2.12}, \ref{Ingredients_V07_z2.12}, \ref{QMaps_V07_z2.12}, \ref{ToomreX}, \ref{VortensityMap}\\
        V07 & $1.13$ & $1.5\times10^{12}$ & $172$ & $1.1\times10^{11}$ & $2.1\times10^{10}$ & $0.15$ & $0.17$ & $0.39$ & $20.4$ & Fig. \ref{QMaps_V07_z1.13} \\
        V08 & $1.33$ & $6.0\times10^{11}$ & $116$ & $1.9\times10^{10}$ & $2.0\times10^{10}$ & $0.42$ & $0.64$ & $0.45$ & $33.1$ & Fig. \ref{QMaps_V08_z1.33} \\
        V13 & $1.78$ & $3.4\times10^{11}$ &  $81$ & $1.2\times10^{10}$ & $1.0\times10^{10}$ & $0.40$ & $0.66$ & $0.46$ & $11.2$ & Fig. \ref{QMaps_V13_z1.78}, \ref{ToomreX}, \ref{VortensityMap} \\
        V19 & $4.88$ & $2.5\times10^{11}$ &  $35$ & $1.3\times10^{10}$ & $7.0\times10^9$   & $0.22$ & $0.42$ & $0.31$ & $24.1$ & Fig. \ref{QMaps_V19_z4.88} \\
        V32 & $3.00$ & $3.0\times10^{11}$ &  $55$ & $1.5\times10^{10}$ & $6.6\times10^9$   & $0.23$ & $0.40$ & $0.39$ & $16.1$ & Fig. \ref{QMaps_V32_z3.00} \\
        \hline
      \end{tabular}
      $$ 
    \end{minipage}
  \end{center}
\end{table*}

We select haloes of virial mass in the range from $10^{11}$ to $10^{12}\Msun$ at $z=1$. For our current analyses, we focus on five galaxies (out of 35) that have a developed disc hosting giant clumps. The properties of the galaxies in our cosmological simulations are listed in Table \ref{CosList}. For one galaxy, V07, we consider two snapshots at different redshifts. For the other galaxies, we examine snapshots in which the discs host giant clumps.

\section{Analysis}
\label{Analysis}
\subsection{Basic Toomre analysis}
\label{ToomreQ}
The Toomre analysis is based on local and liner perturbation theory for axisymmetric density waves. When a razor-thin disc is assumed, a dispersion relation has a quadratic form with respect to wavenumber, and its discriminant is given by
\begin{equation}
  Q=\frac{\sigma\kappa}{AG\Sigma},
  \label{Qsingle}
\end{equation} 
where $\sigma$, $\kappa$ and $\Sigma$ are the radial velocity dispersion, the epicyclic frequency and the surface density, respectively. $G$ is the gravitational constant, and $A$ is a numerical factor (see below). For a gas disc, $\sigma_{\rm gas}$ is given as $\sigma_{\rm gas}^2=c_{\rm s}^2+\sigma_{\rm t}^2$, where $c_{\rm s}$ is the speed of sound, and $\sigma_{\rm t}$ is the radial component of turbulent velocity dispersion. In the case of a highly turbulent disc, such as in a high-redshift galaxy, it can be approximated as $\sigma_{\rm gas}\simeq\sigma_{\rm t}$. For a stellar disc, $\sigma$ is the radial velocity dispersion of the stars. The constant is $A_{\rm gas}=\pi$ for a gas disc and $A_{\rm star}\simeq3.36$ for a stellar disc \citep{t:64,bt:08,e:11}. The similarity of $A$ for the two components implies that the Toomre analyses for gas and stellar discs are almost the same \citep[although see][]{r:01}, and the two disc components should be distinguished according to their kinematic properties, namely by their $\sigma$ and $\kappa$.\footnote{While $\kappa$ is assumed to be the same for gas and stars in a number of cases, we allow different values of $\kappa$ for the two components, to reflect the different rotation curves given the different velocity dispersions (see, \S\ref{howtokappa}).} Toomre $Q$ parameter in Eq. (\ref{Qsingle}) quantifies the balance between the force inwards by self-gravity that is represented by $\Sigma$ and the force outwards by internal pressure and the centrifugal force in rotating frame represented by $\sigma$ and $\kappa$ respectively. The critical value $Q_{\rm crit}=1$ distinguishes between stability and instability; a disc is supposed to be dynamically unstable for axisymmetric perturbations if $Q < Q_{\rm crit}$. 

A disc galaxy consists of gas and stellar components with different kinematics. We analyze such a two-component disc using the approximated formulation of \citet{rw:11}, which is an improved form of the \citet{ws:94} approximation; 
\begin{equation}
  \frac{1}{Q_{\rm 2comp}} =\left \{
  \begin{array}{l}
    \frac{W}{Q_{\rm star}} + \frac{1}{Q_{\rm gas}} \phantom{textte} (Q_{\rm star}>Q_{\rm gas}), \\
    \frac{1}{Q_{\rm star}} + \frac{W}{Q_{\rm gas}} \phantom{textte} (Q_{\rm star}<Q_{\rm gas}),
  \end{array}
  \right.
  \label{Q2comp}
\end{equation} 
where
\begin{equation}
  W = \frac{2\sigma_{\rm star}\sigma_{\rm gas}}{\sigma_{\rm star}^2+\sigma_{\rm gas}^2},
\end{equation} 
and where $Q_{\rm gas}$ and $Q_{\rm star}$ are derived from the quantities of each component separately. From Eq. (\ref{Q2comp}), $Q_{\rm 2comp}$ can be lower than $Q_{\rm crit}$ even if $Q_{\rm gas}$ and $Q_{\rm star}$ are both greater than $Q_{\rm crit}$. \citet{r:01} proposed another formulation for a multi-component $Q$, which we find in Appendix \ref{RafikovQ} to give similar results.

\subsubsection{Deviations from the fiducial $Q$ values}
\label{thicknesscorrection}
 There are several features that may alter $Q$ in the standard Toomre analysis. For example, a non-negligible thickness of a disc naturally weakens the gravitational forces within the disc plane, and thus lowers the value of $Q_{\rm crit}$ for instability \citep[e.g.][]{gl:65,r:92,r:94,wkd:10,bbs:14}. Alternatively, one can compute an approximated thickness-corrected $Q$ parameter keeping the stability criterion at unity as $Q_{\rm thick}=TQ$, where 
\begin{equation}
  T \simeq\left \{
  \begin{array}{l}
    1 + 0.6\left( \frac{\sigma_z}{\sigma_R}\right)^2 \phantom{textttte} (\frac{\sigma_z}{\sigma_R}<0.5), \\
    0.8 + 0.7\frac{\sigma_z}{\sigma_R} \phantom{texttttttte} (\frac{\sigma_z}{\sigma_R}>0.5).
  \end{array}
  \right.
  \label{ThickCorr}
\end{equation} 
The above equation, based on the detailed analysis of \citet{rf:13}, is inferred from analytic calculations presented by \citet[][a top panel of their fig. 3]{r:94}. The correction factor $T$ is computed separately for the gas and the stars, and the $Q_{\rm thick}$ of each of the two components is substituted in Eq. (\ref{Q2comp}). As our standard here, we consider the thin-disc $Q$ as our fiducial value and compare it with the corrected value of $Q_{\rm thick}$ in certain places. 

Another effect that may enhance the instability is a rapid decay of the turbulence by dissipation. This could suppress the support by turbulent pressure in collapsing clumps. \citet{e:11} has estimated that this effect could increase the stability criterion to $Q_{\rm crit}\sim2$ if the gas cooling rate is the same as the dynamical crossing rate, and it would favor the formation of clumps on small scales where turbulent pressure becomes dominant.

It should be noted, however, that these corrections, including other possible corrections, generally need modeling and assumptions which are not always accurate. In our main analysis, we use the $Q$ assuming a razor-thin disc without applying any corrections, and we supplementarily show two-dimensional maps of $Q_{\rm 2comp}$ corrected for the disc thickness according to Eq. (\ref{ThickCorr}) in \S\ref{Qmapsingalaxies} since high-redshift discs have been observed to be significantly thicker than the local spiral galaxies \citep[e.g.][]{ee:06}.


\subsection{Computing Q in the simulations}
\label{treatments}
\subsubsection{Definition of a disc plane}
\label{plane}
The definitions of galactic centres and disc planes follow the method of \citet[][see their Appendix B for details]{mdc:13}. First, the initial guess for the galactic centre is given as the position of the lowest potential. It is then improved iteratively as the centre of mass (COM) of the stellar particles within a radius decreasing from $600$ to $130~{\rm pc}$. 

Next, the rest-frame velocity and the disc plane are iteratively determined based on the angular momentum (AM) vector of the cold gas of temperature $T<1.5\times10^4~{\rm K}$ and stars younger then $100~{\rm Myr}$ in a cylinder of radius $R_{\rm d}$ and height $h_{\rm d}$, and the initial cylinder is of $R_{\rm d}=h_{\rm d}=0.15R_{\rm vir}$. In each iteration, the cylinder is rotated to the coordinates defined by the AM vector within it, and the rest frame is determined by the mass within the cylinder. The cylinder radius $R_{\rm d}$ is updated to contain $85$ per cent of the cold component within a cylinder of radius $R_{\rm d}=0.15R_{\rm vir}$ and hight $h_{\rm d}=1~{\rm kpc}$, and then the height $h_{\rm d}$ is updated to contain $85$ per cent of the cold mass within a cylinder of radius $R_{\rm d}$ and height $h_{\rm d}=R_{\rm d}$. This procedure is repeated until all converge to within five per cent. Throughout the whole procedure, $R_{\rm d}$ and $h_{\rm d}$ are not allowed to fall below $1~{\rm kpc}$ and $0.5~{\rm kpc}$, respectively.

The disc plane can be highly distorted when the galaxy is subject to intense streams including mergers, and the above definition of disc plane could be ambiguous. To avoid this ambiguity, we perform a visual inspection of the disc plane in edge-on density maps. If the disc plane seems to be ill-determined, we do not use that snapshot. For computations of $Q$, we take into account the gas and stars within $|z|<3~{\rm kpc}$ and do not bother to exclude hot gas with $T>1.5\times10^4~{\rm K}$ since the total mass of the hot gas is negligible in comparison with the cold gas. We discuss the robustness of our results with respect to the height considered for the disc in Appendix \ref{Height}. 

\subsubsection{Bulge-star removal}
\label{bulge}
The Toomre analysis is valid for a rotating discs, and it becomes inapplicable for a system that significantly deviates from a disc configuration. Therefore, a galaxy should be decomposed into a disc and a bulge, and the bulge component should be removed (apart from its contribution to $\kappa$, through the potential) for a more accurate determination of $Q$. To identify bulge stars, we adopt a threshold of $J_z/J_{\rm c}<0.7$ for each stellar particle, where $J_z$ and $J_{\rm c}$ are the component of AM parallel to the $z$-axis and the AM of the co-rotating circular orbit with the same orbital energy, respectively \citep{gwm:07,cdb:10}. Thin-disc stars have been observed to have $J_z/J_{\rm c}>0.7$ at the solar neighbourhood in our Galaxy \citep{n:04}. We consider all the gas within $|z|<3~{\rm kpc}$ to be in a disc component, given that the fraction of gas cells that obey the kinematic condition for a bulge is negligible. We examine the impact of including the bulge stars on $Q_{\rm 2comp}$ in Appendix \ref{Bulge}. 

We calculate the stellar bulge-to-total ($B/T$) ratios shown in Table \ref{CosList} using the above threshold. The $B/T$ ratios determined by such a kinematic decomposition are generally greater than those determined by photometric decomposition based on density profile fitting \citep{awn:14}.

\subsubsection{Definition of `Gas' and `Star' components}
\label{twocomponents}
As mentioned in \S\ref{ToomreQ}, the Toomre analysis should be applied separately to \textit{kinematically different} components, rather than simply to gas versus stars. The difference in $\sigma\kappa$ between stars and gas is generally larger than the difference between the constants of $A_{\rm gas,star}=\pi$ and $3.36$ in Eq. (\ref{Qsingle}).\footnote{$A_{\rm star}/A_{\rm gas}=3.36/\pi=1.07$.} For this reason, after removing the bulge stars, we combine the stars younger than $100~{\rm Myr}$ with the gas component since such young stars tend to retain kinematic properties similar to those of the gas. To avoid confusion, we use `Gas' and `Star' to denote the components of gas plus young stars and the remaining disc stars, respectively.

\subsubsection{Smoothing}
\label{smoothing}
If a proto-clump region has a surface density similar to the averaged disc density, then $M_{\rm pc}/M_{\rm d}\sim(l_{\rm pc}/R_{\rm d})^2$, where $M_{\rm pc}$ and $l_{\rm pc}$ are the typical mass and physical scale of the proto-clump region, and $M_{\rm d}$ and $R_{\rm d}$ are the mass and radius of the disc. In this study, our main interest is in the formation of giant clumps with masses of $M_{\rm pc}\sim10^8$--$10^9~\Msun$. The galactic discs in our cosmological simulations typically have masses of $M_{\rm d}\sim10^{10}$--$10^{11}~\Msun$ (see Table \ref{CosList}) and radii of $R_{\rm d}\sim5~{\rm kpc}$. Therefore, we could expect the typical radius of a proto-clump region to be $l_{\rm pc}\sim0.5~{\rm kpc}$. We apply a Gaussian smoothing kernel with this size to all physical quantities that are relevant to the Toomre analyses.

A Gaussian kernel of standard deviation $0.5~{\rm kpc}$ is applied to the gas and the stars, corresponding to a full width at half maximum (FWHM) of $1.2~{\rm kpc}$. Then, we integrate all physical quantities weighted by mass over the vertical range of $|z|<3~{\rm kpc}$ and obtain two-dimensional maps. In \S\ref{SmoothingDependence}, we set the standard deviation to $0.25~{\rm kpc}$ (FWHM$=0.59~{\rm kpc}$) in order to study the dependence on smoothing length.

\subsubsection{How to determine $\kappa$}
\label{howtokappa}
We measure the local $\kappa$ from the actual rotation velocities for the Gas and Star components individually: 
\begin{equation}
  \kappa^2=2\frac{\overline{v_{\phi}}}{R}\left(\frac{\mathrm{d}\overline{v_{\phi}}}{\mathrm{d}R} + \frac{\overline{v_{\phi}}}{R}\right),
  \label{kappa}
\end{equation} 
where $\overline{v_{\phi}}$ is the mass-weighted rotation velocity at $R$. In order to measure a local $\kappa(x,y)$ using Eq. (\ref{kappa}), we have to obtain a local rotation velocity curve $\overline{v_{\phi}}(R)$ at each $\phi$. For this purpose, from the smoothed maps of $\overline{v_{\phi}}(x,y)$ obtained as in \S\ref{smoothing}, we compute $\overline{v_{\phi}}(R)$ in a wedge with an open angle of thirty degrees centred on $\phi$. The obtained $\kappa(R)$ is adopted in the central wedge of two degrees of the wedge. We confirmed that the $\kappa$ computed with this method depends only weakly on the exact open angle used for the wedge.

It should be noted that Eq.(\ref{kappa}) allows $\kappa$ to be an imaginary number. The right-hand-side of Eq. (\ref{kappa}) can become negative where the gas pressure drops steeply since $\overline{v_{\phi}}^2 = R\mathrm{d}\Phi/\mathrm{d}R+(R/\rho)\mathrm{d}p/\mathrm{d}R$ in equilibrium, where $\rho$ and $p$ are the density and the pressure.\footnote{If gas is isothermal and isotropic, $(R/\rho)\mathrm{d}p/\mathrm{d}R\propto\mathrm{d}\log\rho/\mathrm{d}\log R\equiv\gamma$, which is the slope of a density profile. Hence, at a radius of density truncation such as an outer edge of a disc where $\mathrm{d}\gamma/\mathrm{d}R$ has a negatively large value, $\kappa$ tends to be an imaginary number.} A similar phenomenon can occur due to a steep radial gradient of velocity dispersion in a stellar disc. When $\kappa$ is imaginary in either or both of the components, the value of $Q_{\rm 2comp}$ cannot be defined. 

As an alternative, approximated  method, $\kappa$ could be estimated from the circular velocity profile deduced from the potential, namely replacing $\overline{v_{\phi}}$ with $v_{\rm circ}\equiv\sqrt{R\mathrm{d}\Phi/\mathrm{d}R|_{z=0}}$.\footnote{In this case, $\kappa$ is always a real number since $v_{\rm circ}/R\le\kappa\le2v_{\rm circ}/R$ \citep{bt:08}.} In the case of high velocity dispersion, based on the Jeans equation, $\overline{v_{\phi}}$ could be significantly lower than $v_{\rm circ}$, so it could be important to use the more accurate determination based on $\overline{v_{\phi}}$ for high-redshift discs. We note that the epicyclic approximation used in Eq. (\ref{kappa}) is not valid if the pressure gradient term, $(R/\rho)\mathrm{d}p/\mathrm{d}R$, is not negligible compared to the potential gradient one, $R\mathrm{d}\Phi/\mathrm{d}R$ \citep[][]{bt:08}. For the sake of careful analysis, we also show the same computations of $Q_{\rm 2comp}$ but using $v_{\rm circ}$ instead of $\overline{v_{\phi}}$. In Appendix \ref{Spherical} and bottom panels of Fig. \ref{QMaps_V07_z2.12}--\ref{QMaps_V32_z3.00} in \S\ref{Qmapsingalaxies}, we display how our results could change when using $v_{\rm circ}$ to measure $\kappa$, for which the potential is computed assuming spherical symmetry, $v_{\rm circ}\equiv\sqrt{GM(<r)/r}$ where $M(<r)$ is the total mass enclosed within a spherical radius $r$.

\section{Results}
\label{Results}
In this section, we show our results of $Q$ values. We consider the interpretation of $Q$ as follows. A low value of $Q<1$ is consistent with the standard Toomre instability, which is coloured purple in two-dimensional maps of $Q$ in the following Figures, and $Q=1$--$1.8$ (blue in the $Q$ maps) is supposed to be stable according to the linear perturbation theory but could be unstable under non-linear conditions, where perturbations could still grow slowly \citep[e.g.][]{t:64,bt:08}. A value of $Q\sim2$ could be consistent with dissipative instability if the turbulence decays on a timescale similar to the dynamical timescale \citep[][see \S\ref{thicknesscorrection}]{e:11}, and the range with $Q=1.8$--$3.3$ is coloured green in the $Q$ maps. High values of $Q>3.3$ are coloured yellow, red and black  --- these regions are supposed to be stable for the standard Toomre instability. White spots in the $Q$ maps indicate regions where $\kappa$ is imaginary (see \S\ref{howtokappa}).

\subsection{Distribution of $Q$ in clumpy discs}
\label{Qmapsingalaxies}

\begin{figure}
  \includegraphics[width=\hsize]{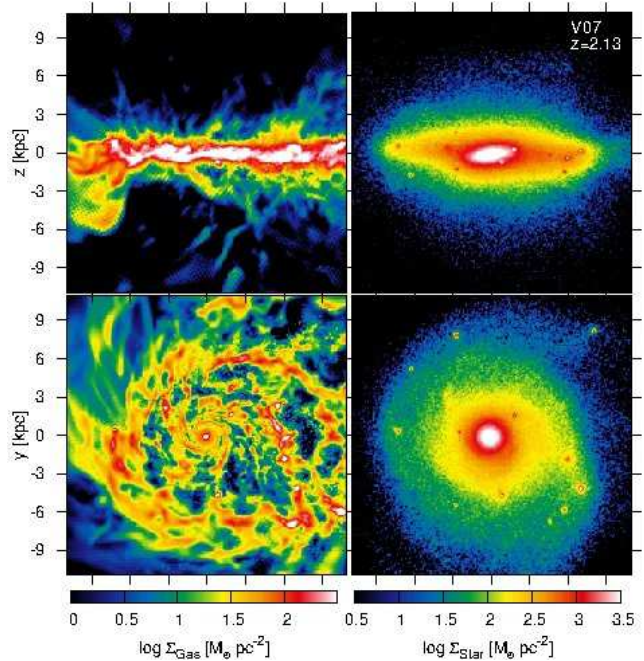}
  \caption{Density maps of V07 at $z=2.13$. The right and left panels show the Gas and Star components, and the top and bottom panels are edge-on and face-on views. No smoothing or bulge removal are applied.}
  \label{DensityMaps_V07_z2.12}
\end{figure}
We apply the analysis described in \S\ref{Analysis} to the simulations presented in \S\ref{Cos-simulation}. First, we look into the clumpy galaxy V07 which has the most massive disc in our sample. The extended disc structure with some giant clumps is present since $z\simeq2.5$. In Fig. \ref{DensityMaps_V07_z2.12}, we show edge-on and face-on density maps of V07 at a relatively high redshift, $z=2.13$, in which clumps can be seen in the Star and Gas components. Young gas-rich clumps are not necessarily visible in the maps of Stars, and old stellar clumps are not necessarily noticeable in the Gas maps. \citet{mdc:13} have found that some of the gas-poor clumps could be `ex-situ' clumps (accreting satellite galaxies), that joined the disc by mergers, while the gas-rich clumps commonly form in-situ in the disc.

\begin{figure*}
  \begin{minipage}{\hsize}
    \includegraphics[width=\hsize]{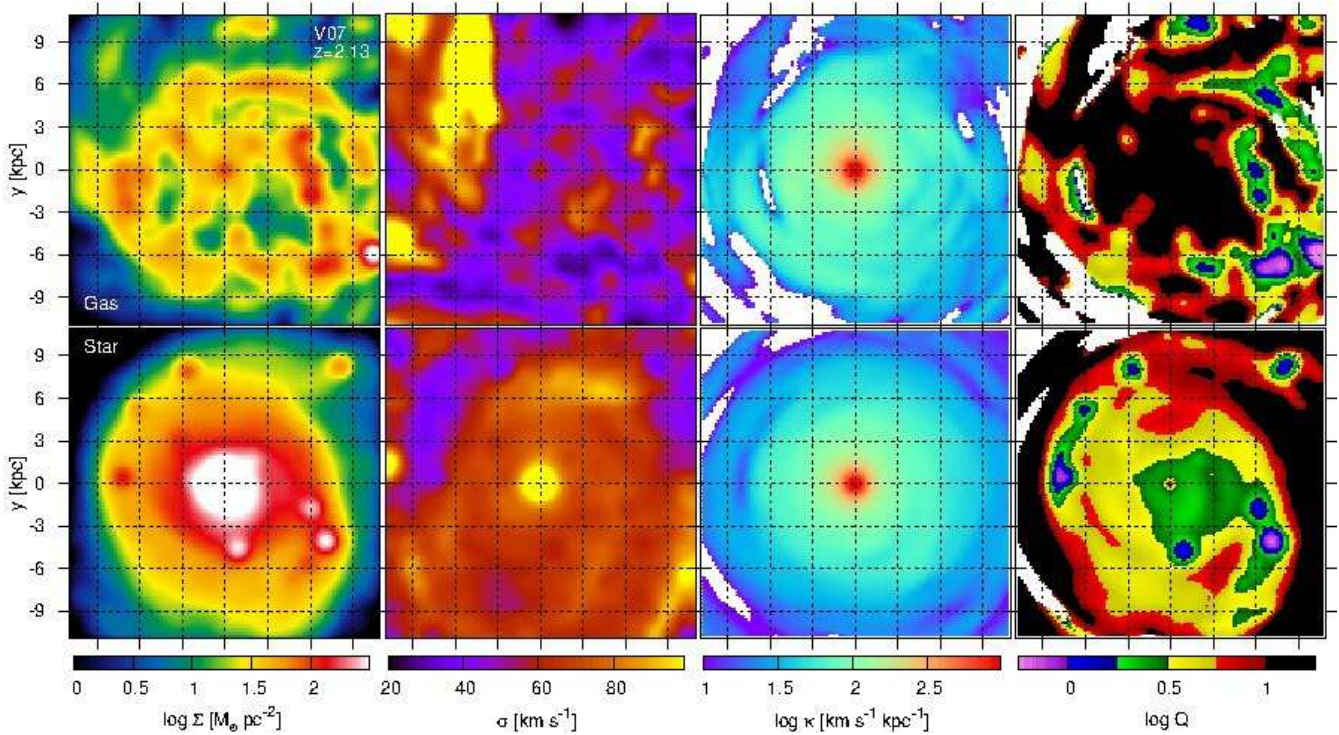}
   \caption{Maps of the ingredients for computing $Q$ values and the resultant $Q$ values of V07 at $z=2.13$. Top and bottom panels indicate the Gas and Star components. From left to right, the panels show surface densities, radial velocity dispersions, epicyclic frequencies and single-component $Q$ values. White colour in the maps of $\kappa$ and $Q$ values indicate regions having imaginary numbers for $\kappa$, where $Q$ values cannot be defined (see \S\ref{howtokappa}). Here, the Gaussian smoothing and the bulge-star removal are applied.}
   \label{Ingredients_V07_z2.12}
  \end{minipage}
\end{figure*}
In Fig.\ref{Ingredients_V07_z2.12}, we show surface density $\Sigma$, radial velocity dispersion $\sigma$, epicyclic frequency $\kappa$ and single-component $Q$ values for the Star and Gas components in V07 at z=2.13 shown in Fig. \ref{DensityMaps_V07_z2.12}. In the disc regions, the extended structures that resemble spiral arms or rings have relatively high $\Sigma_{\rm Gas}$, where clumps are forming along the structures. The velocity dispersion $\sigma_{\rm Gas}$ seems to be relatively low in the regions of high $\Sigma_{\rm Gas}$, presumably because of the dissipative nature of gas. On the other hand, $\sigma_{\rm Star}$ decreases with distance from the centre along with $\Sigma_{\rm Star}$. The epicyclic frequency $\kappa$ is roughly the same between the two components. The rightmost panels show $Q_{\rm Gas}$ (top) and $Q_{\rm Star}$ (bottom). In the map of $Q_{\rm Gas}$, the distribution seems to trace the structure like spiral arms seen in $\Sigma_{\rm Gas}$. The regions with $Q_{\rm Gas}<3.3$ dominate in the middle of the disc ($R\sim5$--$10~{\rm kpc}$), whereas the galactic centre has a high $Q_{\rm Gas}$. Interestingly, the distribution of $Q_{\rm Star}$ is largely different from that of $Q_{\rm Gas}$: the galactic centre has a $Q_{\rm Star}$ lower than in the disc regions except inside and around the stellar clumps. This is because the Star component has a much higher mass-concentration than the Gas even after bulge stars are excluded, although $\sigma_{\rm Star}$ and $\kappa_{\rm Star}$ also peak at the the centre. On the other hand, $\Sigma_{\rm Gas}$ has an extended distribution involving the spiral arms, making $Q_{\rm Gas}$ lower in the middle regions of the disc where $\sigma_{\rm Gas}$ and $\kappa_{\rm Gas}$ are lower than in the central region. None of the components seem to indicate $Q<1$ by itself.

\begin{figure}
  \includegraphics[width=\hsize]{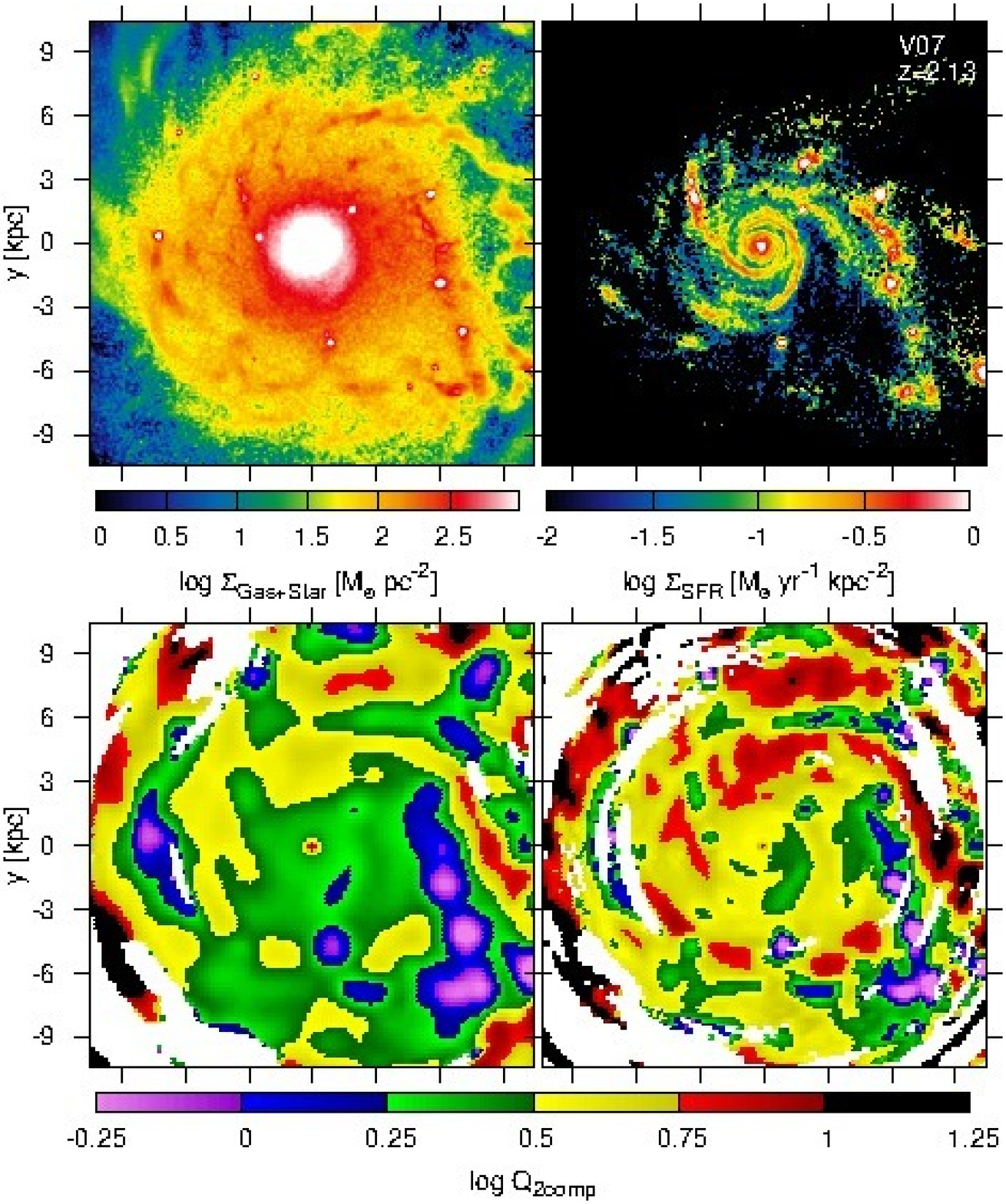}
  \includegraphics[width=\hsize]{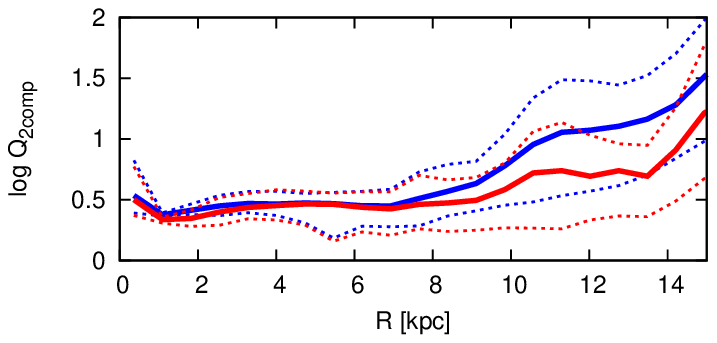}
  \caption{Face-on maps of the total baryon and SFR densities and $Q_{\rm 2comp}$ for V07 galaxy at a redshift of $z=2.13$. The panels shown at the top are the densities of baryons (top left) and SFR (top right). The densities are not smoothed, and bulge stars are not removed. The panels shown below are the two-component $Q$ values smoothed on a $\sim 1~{\rm kpc}$ scale; $Q_{\rm 2comp}$ without the thickness correction (middle left) can be compared to the one corrected for disc thickness (middle right). The bottom panel refers to the profiles of annularly averaged $Q_{\rm 2comp}$, in the cases where $\kappa$ is determined from $\overline{v_{\phi}}$ (red solid) and $v_{\rm circ}\equiv\sqrt{GM(<r)/r}$ (blue solid). The dashed lines refer to the $1$-$\sigma$ deviations from the averaged values in each radial bin.}
  \label{QMaps_V07_z2.12}
\end{figure}
The top panels of Fig. \ref{QMaps_V07_z2.12} show the maps of surface density of baryons (left) and star-formation rate (SFR, right). The SFR is measured from stars younger than $100~{\rm Myr}$,\footnote{H$\alpha$ emittion line is sensitive only to ages $\lsim10~{\rm Myr}$, and rest-frame ultra-violet light traces stars the ages of which are $\sim10$--$100~{\rm Myr}$ while optical colours are sensitive up to ages of $\sim1~{\rm Gyr}$. In this sense, SFRs defined in this paper would be consistent the best with ultra-violet observations.}. The star-formation activity is concentrated in massive clumps. In the SFR map, the galaxy appears more clumpy than in the baryon density map. It should be noted that the bulge appears less pronounced in the SFR map because its gas has already been depleted. We expect the SFR maps to match better with observations based on H$\alpha$ or blue colors than the surface density maps. The middle row shows the maps of $Q_{\rm 2comp}$ with the razor-thin model (left) and the same but corrected for thickness (right, see \S\ref{thicknesscorrection}). In the Figure, the regions with $Q_{\rm 2comp}<1$ (purple) can only be seen inside and around the clumps. The disc (inter-clump) regions generally have $Q_{\rm 2comp}\gsim1.8$. In the bottom panel, we show radial profiles of annularly averaged $Q_{\rm 2comp}$. Our fiducial $Q$ values where $\kappa$ is derived from $\overline{v_{\phi}}$ (solid) is compared with $Q_{\rm 2comp}$ where $\kappa$ is approximated using $v_{\rm circ}$ (see \S\ref{howtokappa}). The averaged profiles indicate $Q_{\rm 2comp}\gsim1.8$ at all radii, indicating linear Toomre stability, but we will see that new clumps are still forming in the disc of this galaxy in \S\ref{ThinSlices}. 

\begin{figure}
  \includegraphics[width=\hsize]{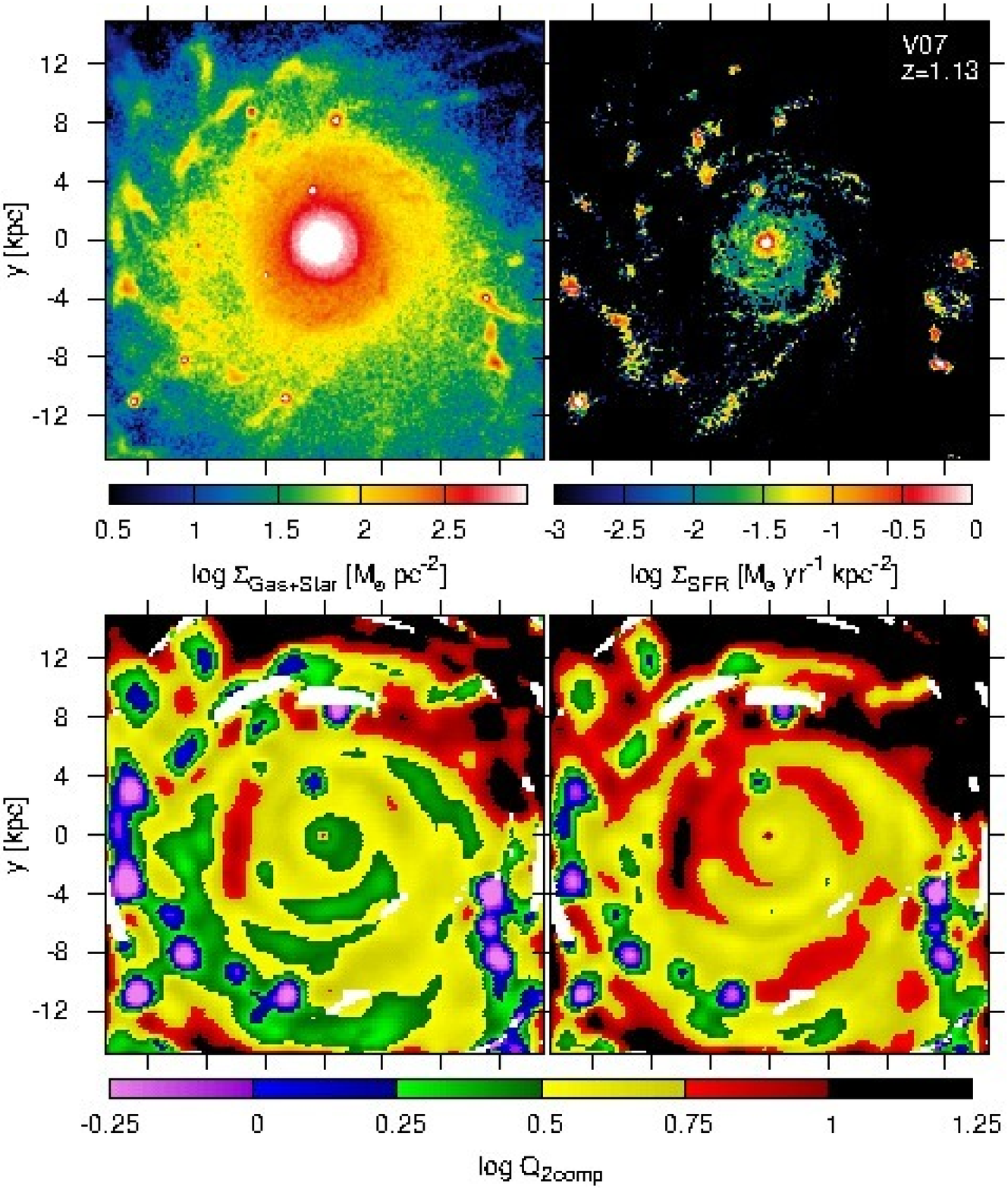}
  \includegraphics[width=\hsize]{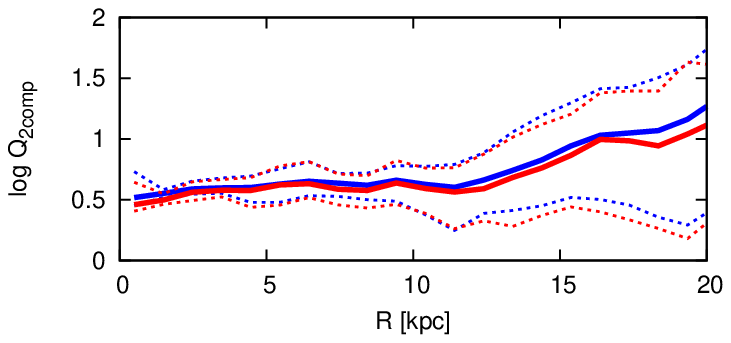}
  \caption{Same as Fig. \ref{QMaps_V07_z2.12} but for V07 at $z=1.13$.}
  \label{QMaps_V07_z1.13}
\end{figure}
\begin{figure}
  \includegraphics[width=\hsize]{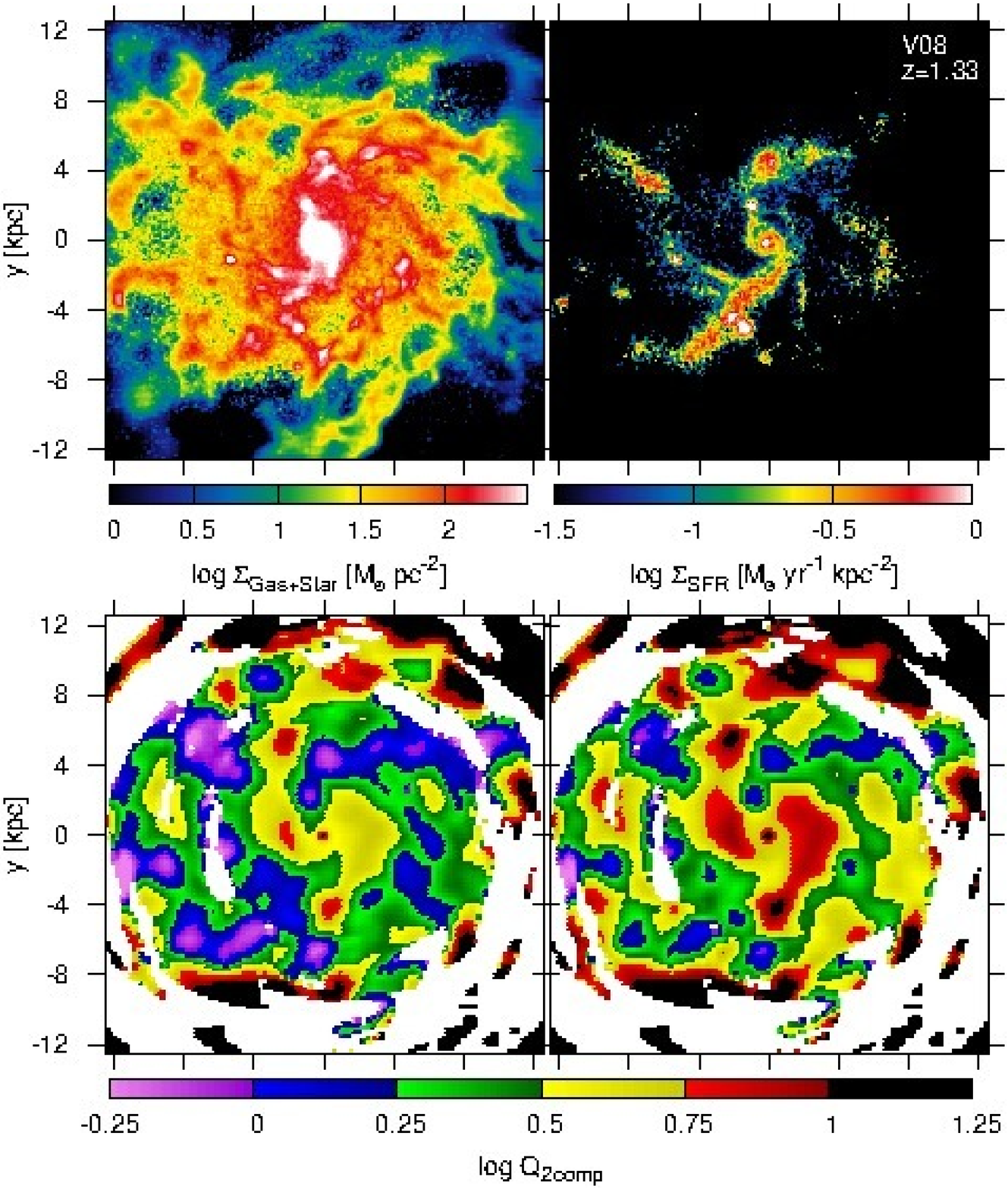}
  \includegraphics[width=\hsize]{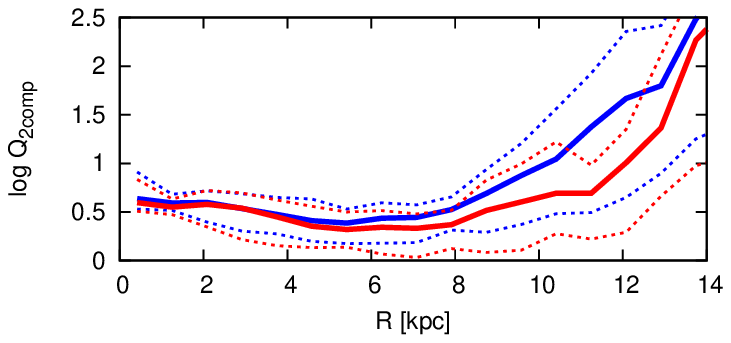}
  \caption{Same as Fig. \ref{QMaps_V07_z2.12} but for V08 at $z=1.33$.}
  \label{QMaps_V08_z1.33}
\end{figure}
\begin{figure}
  \includegraphics[width=\hsize]{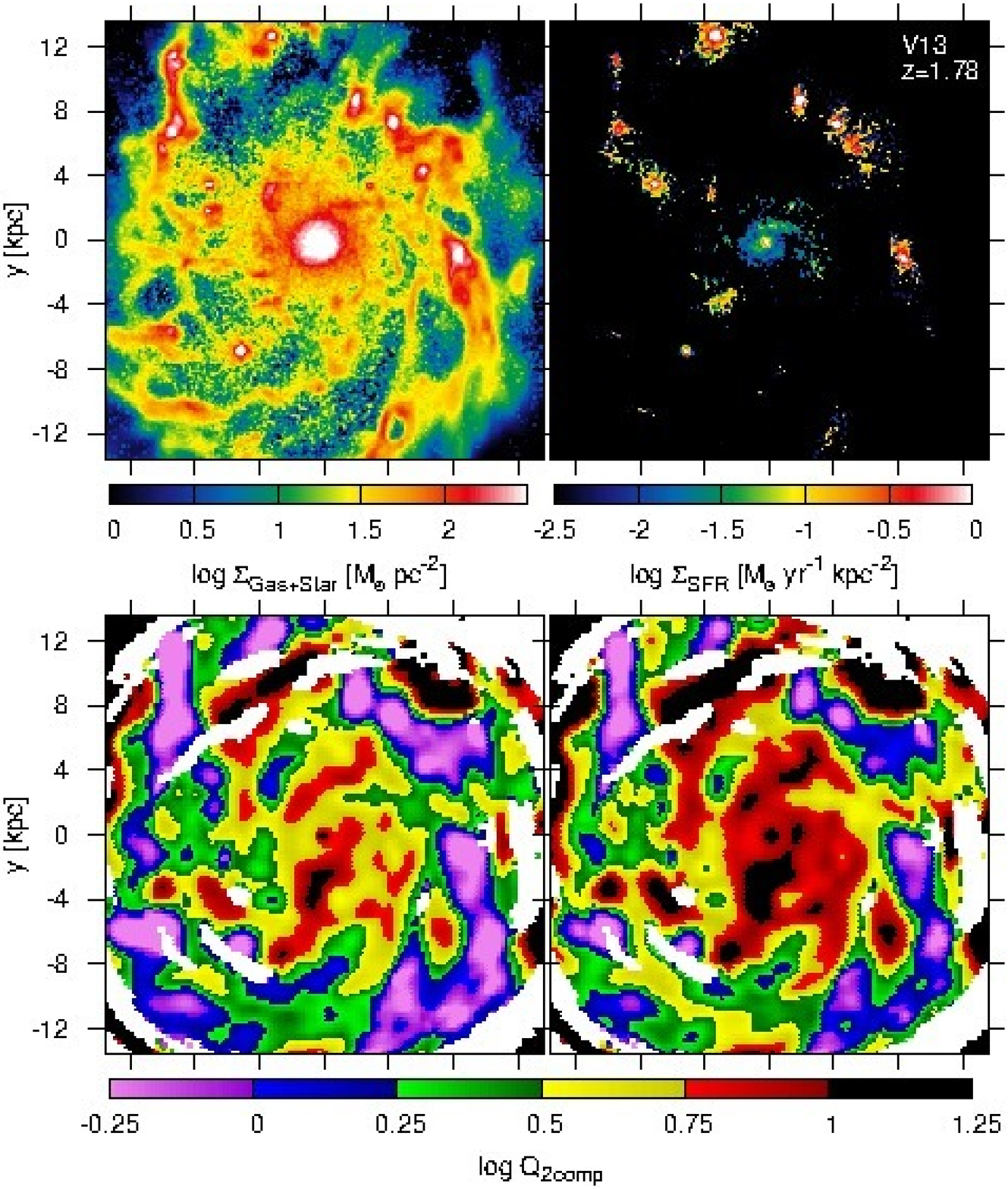}
  \includegraphics[width=\hsize]{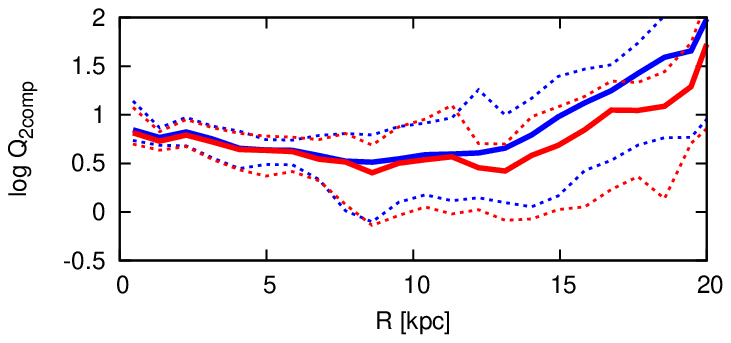}
  \caption{Same as Fig. \ref{QMaps_V07_z2.12} but for V13 at $z=1.78$.}
  \label{QMaps_V13_z1.78}
\end{figure}
\begin{figure}
  \includegraphics[width=\hsize]{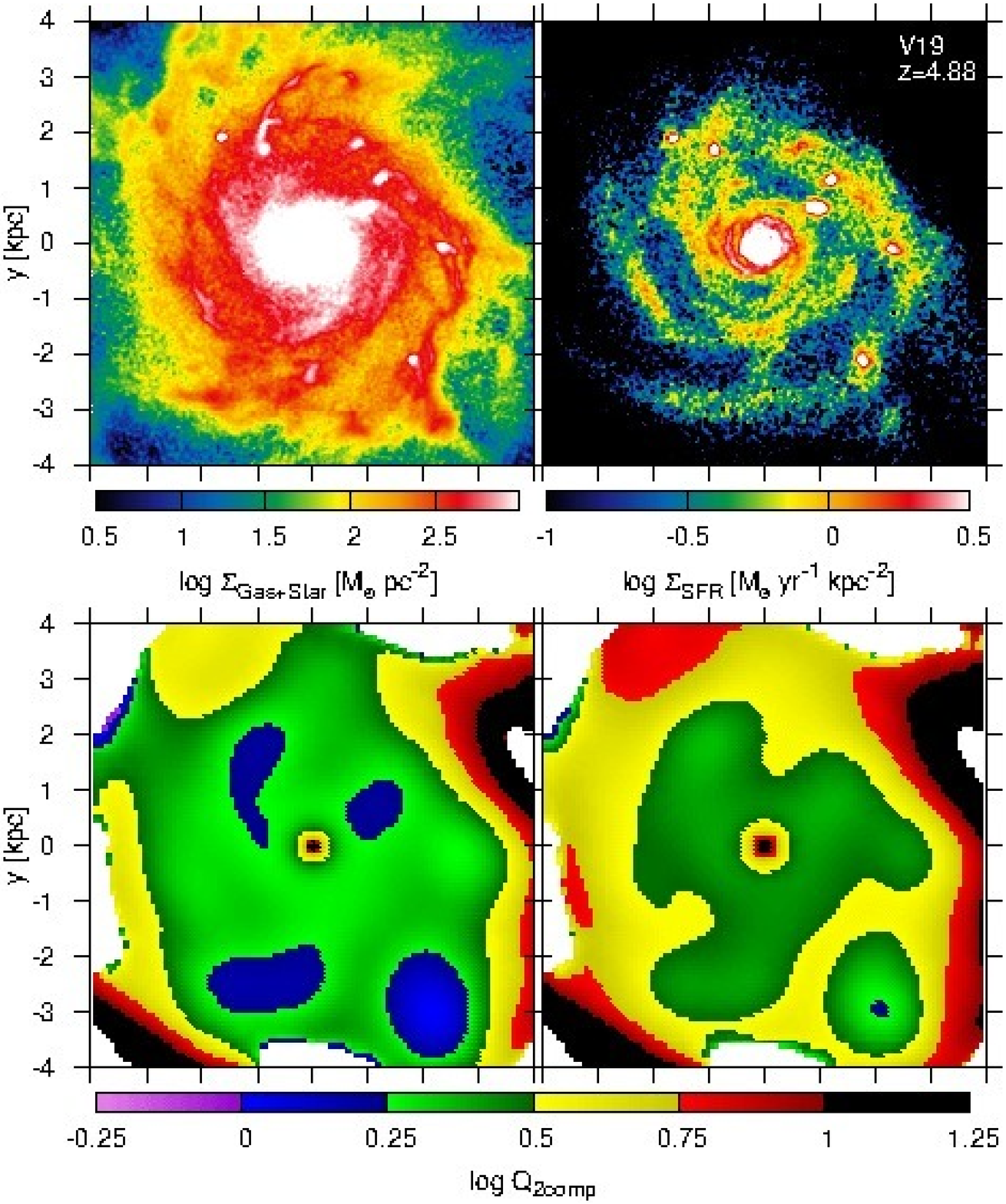}
  \includegraphics[width=\hsize]{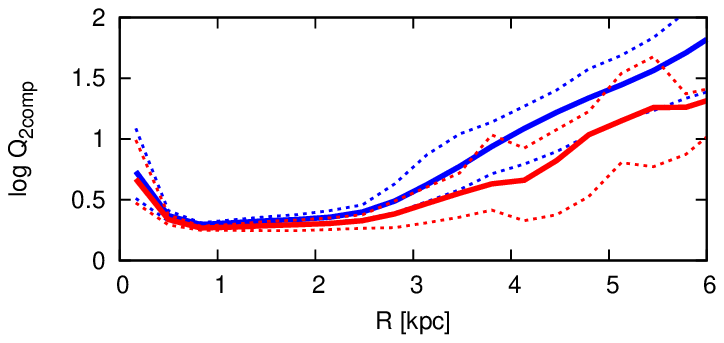}
  \caption{Same as Fig. \ref{QMaps_V07_z2.12} but for V19 at $z=4.88$.}
  \label{QMaps_V19_z4.88}
\end{figure}
\begin{figure}
  \includegraphics[width=\hsize]{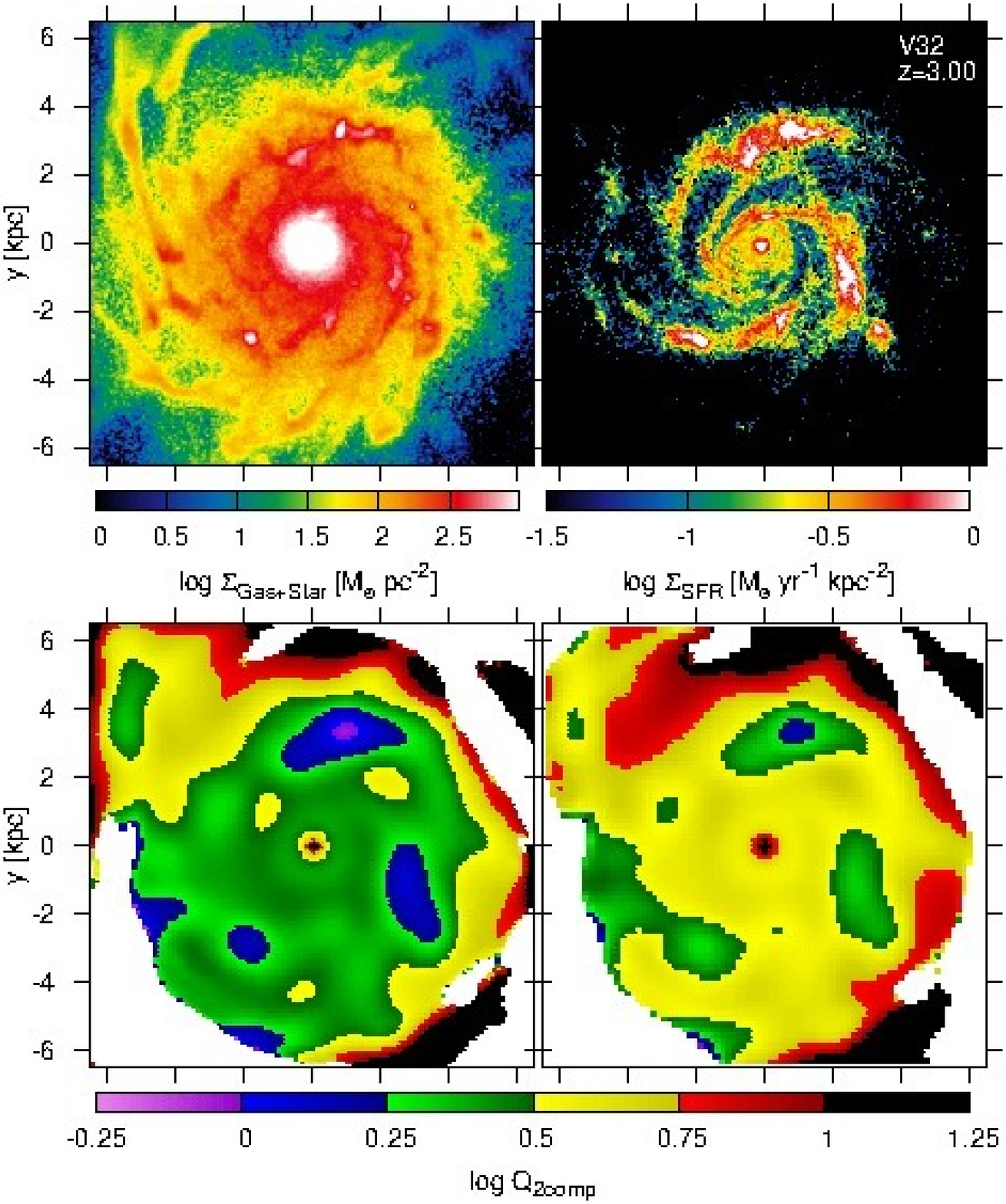}
  \includegraphics[width=\hsize]{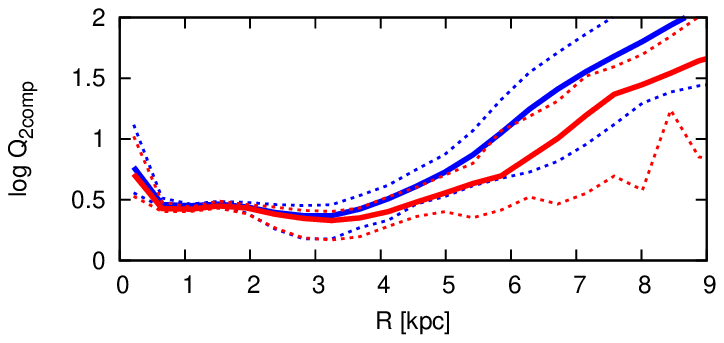}
  \caption{Same as Fig. \ref{QMaps_V07_z2.12} but for V32 galaxy at a redshift of $z=2.85$.}
  \label{QMaps_V32_z3.00}
\end{figure}
The clumpy phase of V07 lasts long, until $z\sim0.8$ at least.\footnote{This galaxy was run until $z\simeq0.8$, but it may remain clumpy to lower redshifts.} In Fig. \ref{QMaps_V07_z1.13}, we show our Toomre analysis for V07 at a relatively low redshift, $z=1.13$, where the mass fraction of the Gas component is only $f_{\rm Gas}^{\rm disc}=0.17$ compared to $0.37$ in the snapshot at $z=2.13$ shown in Figs. \ref{DensityMaps_V07_z2.12}--\ref{QMaps_V07_z2.12} (see Table \ref{CosList}). The galaxy still hosts giant clumps in the disc. The maps of $Q_{\rm 2comp}$ indicate in general higher values than in the earlier snapshot. Values of $Q_{\rm 2comp}\lsim1.8$ (purple and blue) can only be seen inside and around the clumps, and the inter-clump regions generally have $Q_{\rm 2comp}\gsim1.8$.

We next refer to the other galaxies listed in Table \ref{CosList}. Fig. \ref{QMaps_V08_z1.33} shows the results for V08 at $z=1.33$. The virial mass of this galaxy is about half that of V07 at the similar redshift, and it develops a clumpy disc emerging at $z\simeq1.4$. The disc Gas fraction of this galaxy at this redshift is $f_{\rm Gas}^{\rm disc}=0.64$, significantly higher than that of V07 (see Table \ref{CosList}), and the clumps are generally young and gas-rich. As in V07, however, $Q_{\rm 2comp}$ is only lower than unity inside and around the clumps while the inter-clump regions seem to show values of $Q_{\rm 2comp}$ significantly higher than unity.

Fig. \ref{QMaps_V13_z1.78} shows the results for V13 at $z=1.78$. The disc of V13 has a high gas fraction, similar to V08. The map of $Q_{\rm 2comp}$ shows a ring-like feature at $R\sim5$--$10~{\rm kpc}$, which consists of large patches of relatively low $Q_{\rm 2comp}$. These patches still correlate with the compact clumps although they are more extended than in the other galaxies. Some of the inter-clump regions show $Q_{\rm 2comp}\gsim1.8$, as in the other galaxies.

Cosmological simulations indicate that galaxies can form small and massive discs at very early epochs. This is seen in our simulations, and also in \citet{fdc:15}, who have estimated that small discs account for nearly seventy per cent of the galaxies of $M_{\rm star}>10^{10}~{\rm M_\odot}$ at $z=8$. Fig. \ref{QMaps_V19_z4.88} and \ref{QMaps_V32_z3.00} show the results for two of our high-redshift small discs, V19 at $z=4.88$ and V32 at $z=3.00$. Several clumps can be seen in the maps of surface density and SFR. However, these galaxies indicate $Q_{\rm 2comp}\gsim1.8$ in their inter-clump regions, similar to the clumpy galaxies at lower redshifts in our simulations, although the high-redshift discs appear to have slightly lower $Q_{\rm 2comp}$ than the other low-redshift ones.

The smoothing of FWHM$=1.2~{\rm kpc}$ may be too large for the compact discs of V19 and V32 (see \S\ref{smoothing}). In Appendix \ref{SmoothingDependence}, we show the results for all galaxies listed in Table \ref{CosList} but computed with a smaller smoothing size of FWHM$=0.59~{\rm kpc}$. We confirm that, even with the smaller smoothing length, the maps indicate $Q_{\rm 2comp}$ significantly higher than unity in the inter-clump regions.

\subsection{The value of $Q$ in proto-clump regions}
\label{ThinSlices}

Despite the high values of $Q_{\rm 2comp}$ in the inter-clump regions of our simulated discs, giant clumps continue to form there. This is evident from the fact that the clumpy phases typically last for $\gsim10^9~{\rm yr}$, while the timescale for clump migration to the disc centre is only $\sim10^8~{\rm yr}$ \citep[e.g.][]{bee:07,bpr:14,ebe:08,dsc:09,cdb:10}.

\begin{figure}
  \includegraphics[width=\hsize]{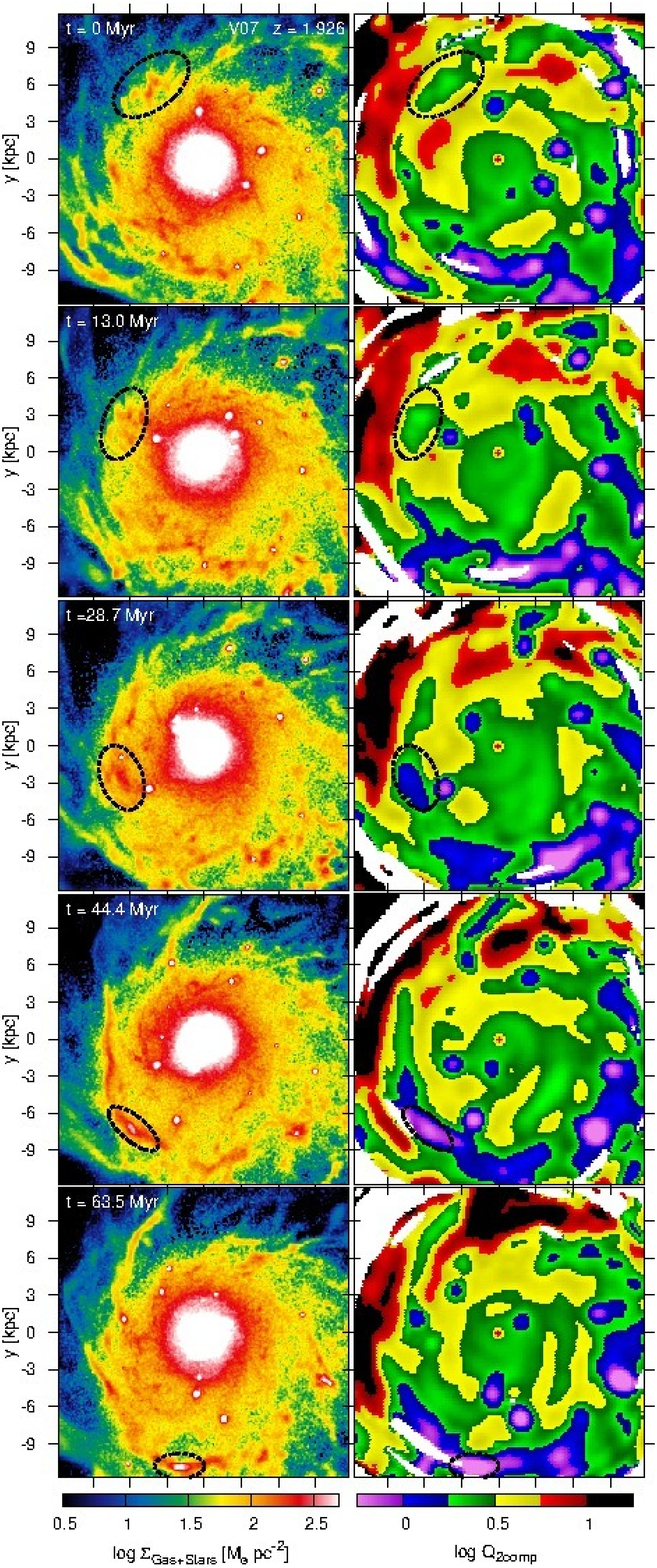}
  \caption{Maps of baryon surface densities (left) and $Q_{\rm 2comp}$ without the thickness correction (right) in the run of V07 during formation of a new clump. A proto-clump region is marked with black dashed circles. Times elapsed since the topmost panels ($z=1.926$) are indicated on the top right corners of the left panels.}
  \label{ProtoClumpCase1}
\end{figure}
\begin{figure}
  \includegraphics[width=\hsize]{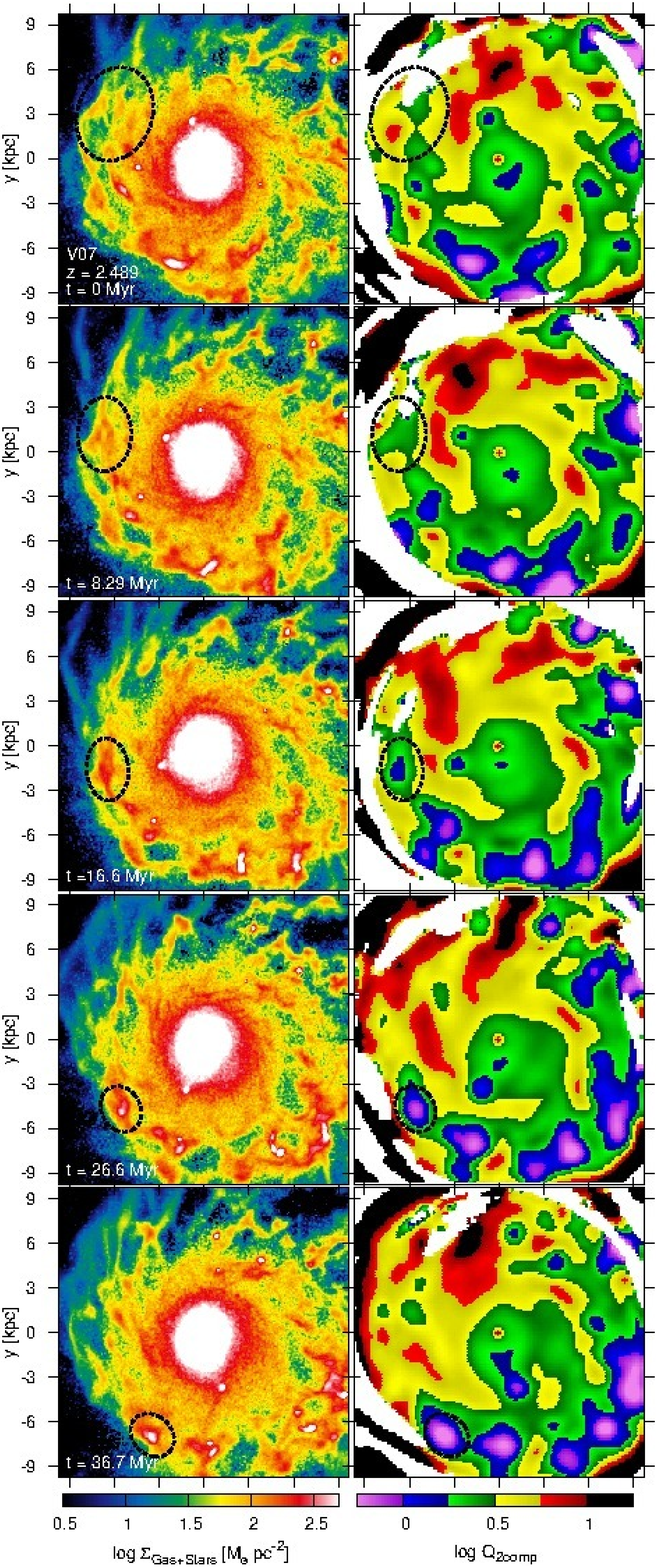}
  \caption{The same as Fig. \ref{ProtoClumpCase1} but for another case of clump formation in V07. A proto-clump region is marked with black dashed circles. Times elapsed since the topmost panels ($z=2.489$) are indicated on the bottom right corners of the left panels.}
  \label{ProtoClumpCase2}
\end{figure}

In what follows, we show that formation of giant clumps can start in regions where $Q_{\rm 2comp}$ is significantly greater than unity. In Fig. \ref{ProtoClumpCase1} and \ref{ProtoClumpCase2}, we show the frame-by-frame advance of surface density and $Q_{\rm 2comp}$ in short periods during the evolution of the simulated galaxy, where apparently weakly perturbed `proto-clump' Lagrangian regions develop into compact massive clumps. Fig. \ref{ProtoClumpCase1} shows V07 starting at $z=1.926$, and the black dashed ellipses in the five panels from top to bottom chase a proto-clump region during a period of $63.5~{\rm Myr}$. In the top and second-from-top panels, the proto-clump is seen as a density fluctuation with a small amplitude in which $Q_{\rm 2comp}=1.8$--$3.3$ (green) and larger when the thickness correction is applied. In the middle panels, at $28.7~{\rm Myr}$ after the starting top panels, the surface density of the forming clump increases and $Q_{\rm 2comp}$ decreases accordingly to $1.0$--$1.8$ (blue). In the following snapshots, a compact core with high density appears at the centre of the proto-clump region, forming a giant clump there in which $Q_{\rm 2comp}<1$ (purple). Fig. \ref{ProtoClumpCase2} shows a similar sequence of snapshots for V07 starting at $z=2.489$, which also demonstrates clump formation starting with a high value of $Q_{\rm 2comp}$ in the weakly perturbed proto-clump region. In the left of the top panels, several gas clouds with low density contrast are seen within the black dashed ellipse, and it is difficult to see which cloud evolves into the massive compact clump. In the second panels, the gas clouds seem to merge with each other, while $Q_{\rm 2comp}$ is still $1.8$--$3.3$ (green). In the middle panels, the density of the proto-clump region becomes high and $Q_{\rm 2comp}=1.0$--$1.8$ (blue). Then, in the following panels, when a massive compact clump forms in the high density gas cloud, $Q_{\rm 2comp}$ reaches a value below unity (purple). In these cases, the formation of giant clumps does not start in a Toomre unstable state with $Q_{\rm 2comp}$ lower than or near unity.

To further investigate the value of $Q_{\rm 2comp}$ in the proto-clump regions where clumps start to develop, we perform automatic clump identification following the method described in \citet{mdc:13}. First, using a cloud-in-cell interpolation, we deposit mass in a uniform grid with a cell size of $70~{\rm pc}$. The procedure washes out noise at the simulation resolution level but retains the desired clumpy structures. Then, we apply a spherical Gaussian smoothing with FWHM$={\rm min}(2.5~{\rm kpc},0.5R_{\rm d})$ to the gas density field given on the uniform grids. The Gaussian filter smooths out large-scale structures comparable to the disc scale. Then, we obtain the residual of the two densities: $\delta\equiv(\rho_{\rm N}-\rho_{\rm W})/\rho_{\rm W}$, where $\rho_{\rm N}$ and $\rho_{\rm W}$ are the unsmoothed and smoothed densities. Next, we group neighbouring gas cells having $\delta>10$ into a clump. Stellar particles in a clump are assigned to the gas cells that contain the stars. Then, we calculate the total baryon mass of the clump $M_{\rm cl}$. The position of the clump is defined as the centre of the gas cell having the highest $\rho_{\rm N}$, and the physical size of the clump $l_{\rm cl}$ is defined as the radius of a sphere with the same volume as the clump: $(4\pi/3)l_{\rm cl}^3=N\times(70~{\rm pc})^3$, where $N$ is the total number of grid cells within the clump. By assuming a uniform density within the clump, we estimate the dynamical crossing time of the newly formed clump as $t_{\rm dyn, cl}\equiv\pi/2\sqrt{l_{\rm cl}^3/(GM_{\rm cl})}$.

By applying the clump identification to our simulation snapshots separated by time-intervals of $1$--$10~{\rm Myr}$, the clump can be traced along its orbit back to the position and time when the clump forms; the formation position and time are defined in the earliest snapshot where the clump is detected for the first time. The proto-clump position is inferred in the snapshot just prior to the formation snapshot by linearly interpolating the clump orbit back in time using its position and velocity at formation.\footnote{The orbital extrapolation from the formation position to the proto-clump position is performed using a velocity vector in the cylindrical coordinates: $\mathbf{v}=(v_{R},v_{\phi},v_{z})$.}

We then compute $Q_{\rm 2comp}$ at the proto-clump position. To focus on the formation of massive clumps, we exclude clumps that have $M_{\rm cl}<10^8~{\rm M_{\odot}}$ when they are identified for the first time. We also exclude the transient clumps whose total lifetimes until they disappear are shorter than their internal dynamical times $t_{\rm dyn, cl}$ measured at their formation. The dynamical times of the new massive clumps are typically on the order of $\sim10~{\rm Myr}$. Our analysis also excludes clumps forming outside the disc, at $r>0.15R_{\rm vir}$; these are likely to be ex-situ clumps.

\begin{figure}
  \includegraphics[width=\hsize]{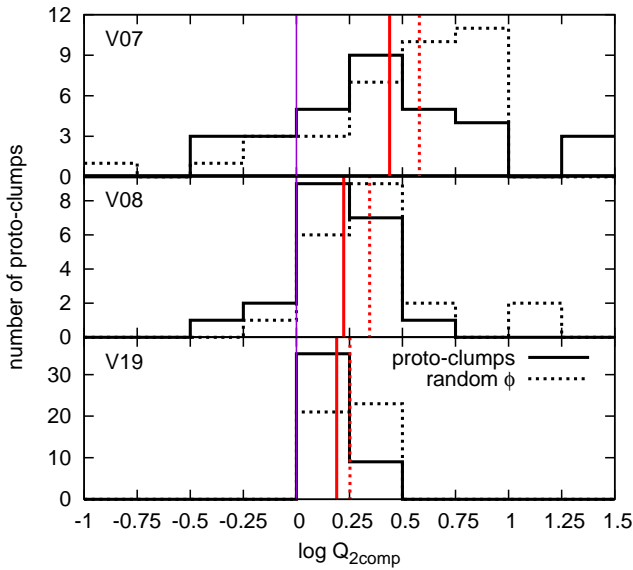}
  \caption{The distribution of $Q_{\rm 2comp}$ at the proto-clump positions (solid histograms) in the clumpy phases of V07 (top), V08 (middle) and V19 (bottom). Shown in comparison is the distribution of $Q_{\rm 2comp}$ in inter-proto-clump positions (dashed histograms), where the angular positions in the discs $\phi$ are chosen at random avoiding the 1-${\rm kpc}$ neighbourhood of the proto-clumps without varying the radial positions $R$. The vertical purple lines indicate $Q_{\rm 2comp}=1$, and the vertical red lines indicate the median values of $Q_{\rm 2comp}$ at the proto-clumps (red solid) and the inter-proto-clumps (red dashed). We find that most of the proto-clumps have $Q_{\rm 2comp}>1$, and that a significant fraction have $Q_{\rm 2comp}>1.8$ and even $Q_{\rm 2comp}>3.3$ (especially in V07). The inter-proto-clump regions typically have $Q_{\rm 2comp}$ higher than the proto-clump regions.}
  \label{Qhist}
\end{figure}
We perform the above analyses for V07, V08 and V19 during their clumpy disc phases: from $z\simeq2.48$ to $0.83$ in V07, from $z\simeq1.50$ to $0.89$ in V08 and from $z\simeq5.67$ to $2.57$ in V19. During these periods, we obtain 37, 20 and 44 proto-clumps for V07, V08 and V19, respectively. In V07, five of the 37 proto-clump regions have imaginary $\kappa$ at their centres, and they are excluded from the analysis, while there are no imaginary $\kappa$ in any of the proto-clumps in the other two galaxies. We show the distribution of $Q_{\rm 2comp}$ at the extrapolated proto-clump positions in Fig. \ref{Qhist} (solid histograms). We first notice that most of the proto-clumps have $Q_{\rm 2comp}>1$. The median values of $Q_{\rm 2comp}$ on the proto-clump positions are $2.75$, $1.67$ and $1.55$ for V07, V08 and V19, respectively, significantly higher than unity in all cases. While in V19 most of the proto-clumps show $Q_{\rm 2comp}=1$--$1.8$,\footnote{The time-intervals between the snapshots are $1$--$5~{\rm Myr}$ in V19 while $5$--$15~{\rm Myr}$ in V07 and V08. Although the time-intervals in V07 and V08 are comparable to $t_{\rm dyn, cl}$, those in V19 could be shorter than the typical value of $t_{\rm dyn, cl}$. Hence, the proto-clump regions in V19 are closer to their post-formation phases than those in V07 and V08, which generally indicate lower $Q_{\rm 2comp}$ than those in their proto-clump phases. In Fig. \ref{Qhist}, the median values of $Q_{\rm 2comp}$ in the proto-clump regions in V19 may be biased to a lower value because of the shorter time-intervals between the snapshots.} a significant fraction of the V08 proto-clumps have $Q_{\rm 2comp}=1.8$--$3.3$, and $38$ per cent of the proto-clumps in V07 have $Q_{\rm 2comp}>3.3$, with some proto-clumps showing $Q_{\rm 2comp}\sim10$. These high values of $Q_{\rm 2comp}$ on the proto-clumps quantify our results from the examples shown in Fig. \ref{ProtoClumpCase1} and \ref{ProtoClumpCase2}. The lower 90th percentiles of $Q_{\rm 2comp}$ at the proto-clump positions are $18.2$, $2.69$ and $1.87$ for V07, V08 and V19, respectively. These values may serve as upper limits of $Q_{\rm 2comp}$ for clump formation. We find that a large fraction of the proto-clump regions in our simulated galaxies in VDI at high redshifts show values of $Q_{\rm 2comp}\gsim2$ on 1-${\rm kpc}$ scale for massive clumps of $M_{\rm cl}\gsim10^8~{\rm M_{\odot}}$, which are inconsistent with the standard Toomre instability theory. 

In order to compare to the values of $Q_{\rm 2comp}$ in the inter-proto-clump regions, the dotted lines in Fig. \ref{Qhist} indicate the distribution of $Q_{\rm 2comp}$ at positions where the angular positions $\phi$ of the proto-clumps are replaced with random angles without varying $R$, avoiding regions within $1~{\rm kpc}$ from the original and other proto-clump positions. The Figure shows that $Q_{\rm 2comp}$ at the inter-proto-clump region tends to be higher than that at the proto-clump position; the median values are $3.79$, $2.21$ and $1.79$ for V07, V08 and V19, respectively. We learn that the value of $Q_{\rm 2comp}$ is still a good indicator of instability, in the sense that it is anti-correlated with clump formation, although the critical value is higher than unity, typically $Q_{\rm 2comp}\sim2$--$3$ (even before we apply a thickness correction).

\begin{figure}
  \includegraphics[width=\hsize]{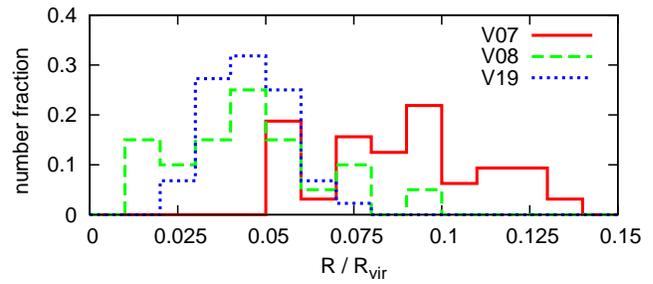}
  \caption{Histograms of galactocentric distances of the proto-clumps. The distances are in units of the virial radii at the times when the clumps form.}
  \label{Rhist}
\end{figure}
Fig. \ref{Rhist} shows the distribution of the galactocentric distances of the proto-clumps in V07, V08 and V19. The histogram of V07 (red) is skewed towards higher values of $R/R_{\rm vir}$ reflecting the fact that this galaxy has a very extended disc. Generally, massive clumps form in regions from the middle to the outer edge of the disc. We do not find a clear correlation between $Q_{\rm 2comp}$ at the proto-clump positions and their galactocentric distances.

\section{Discussion}
\label{Discussion}
As we showed in \S\ref{Results}, non-negligible fractions of the proto-clump regions in high-redshift disc galaxies have $Q$ values significantly higher than unity. This result implies that clump formation may not be simply interpreted in terms of standard linear Toomre instability with $Q_{\rm 2comp}<1$. In the following subsections from \S\ref{Possibility1} to \S\ref{Possibility3}, we discuss possible interpretations of this result and the physical mechanisms of clump formation. In \S\ref{Observations}, we compare our results with and observational estimations of $Q$ for disc galaxies at high- and low-redshifts. 

\subsection{Inaccuracy of the Toomre criterion and the measurements of $Q$}
\label{Possibility1}
Clump formation could take place by Toomre instability, but the criterion of $Q=1$ and/or the measurements of $Q_{\rm 2comp}$ might be inaccurate in realistic non-linear situations such as in cosmological simulations. For example, it should be noted that the Toomre analysis refers to axisymmetric perturbation, while the perturbations in the cosmological simulations could significantly deviate from the axial symmetry. Moreover, in the Toomre analysis, the dispersion relation of a stellar disc assumes the axisymmetric Schwartzchild distribution function \citep{t:64,r:01,bt:08}, while in real galaxies the density and velocity distribution can deviate from the assumed distribution function. 

\citet{bbs:14} have performed a simulation of an isolated, isothermal gas disc and compared the results with their Toomre analysis taking the disc thickness into account. Their results indicated good agreement between the analytic predictions concerning the annular instability and the simulation in which the initial conditions were tailored for linear perturbations. In their simulation, the annular density fluctuations with $Q<1$ are narrow enough to break up into clumps. One can confirm from their results that the Toomre analysis provides sensible prediction for the behaviour of axisymmetric linear perturbations in isolated, uniform discs.

We note that the two-component $Q$ that we have used provides values lower than what one would have obtained with a single-component analysis. However, the decomposition into Gas and Stars might be an over-simplification. While we have used a reliable approximation for a two-component Toomre analysis, the simulated Star and Gas discs may actually be composed of a larger number of components that are kinematically distinguishable, which may be better treated with a more sophisticated multi-component analysis.\footnote{We confirmed, though, that our results are robust to exclusion of hot gas with $T > 1.5 \times 10^4~{\rm K}$.} 

In Appendix \ref{App}, we examine the robustness of our results with respect to certain choices made in our analysis: the formulation of $Q_{\rm 2comp}$, how to measure $\kappa$, the bulge-star removal, the vertical cut-off for discs and the smoothing scale. We find that our results are only weakly sensitive to these choices, and in particular that other choices do not yield significantly lower values of $Q_{\rm 2comp}$. If giant clumps in our simulations really formed with $Q_{\rm 2comp}<1$, we would have overestimated the $Q$ values by a factor of two or more. However, we do not find such large uncertainties in the measurements of $\Sigma$, $\sigma$ or $\kappa$.

\subsection{Non-linear effects and small-scale Toomre instability}
\label{Possibility2}
The Toomre analysis is based on local and linear perturbation theory. However, additional effects such as higher-order and global instability could play an important role in forming clumps. Although the instability criterion is originally supposed to be $Q<Q_{\rm crit}=1$ for a razor-thin disc in the local linear perturbation theory of \citet{t:64}, $Q_{\rm crit}$ can depend on various effects such as non-linear and global growth of perturbation, disc thickness, rapid decay of turbulence, external potential field and other perturbances including mergers. Moreover, the value of $Q$ can depend on physical scale, and the numerical resolution in the simulations is also a relevant issue. 

It has been proposed that discs in collisionless simulations are unstable up to $Q_{\rm crit}=1.7$ due to non-linear and global instability \citep[e.g.][]{bt:08,fbs:11,mk:14}. \citet{hs:15} proposed $Q_{\rm crit}=1.3$ using $N$-body simulations with resolutions higher than the other studies. 


Rapid decay of turbulence, possibly associated with gas dissipation, could destabilize the disc on small scales \citep{e:11}. If the turbulence decays on a timescale shorter than the crossing time within the proto-clump, the turbulent pressure cannot support gravitational collapse. In this case, the instability criterion may rise to $Q_{\rm crit}\sim2$ or maybe even higher. The resulting small clumps may grow by non-linear effects such as mergers with other clumps and further accretion. On the other hand, we recall that the the disc thickness tends to stabilize the disc (\S\ref{thicknesscorrection}). \citet{gl:65} showed that the thickness effect effectively decreases $Q_{\rm crit}$ to $\simeq0.67$ (with $Q$ still being computed in the thin-disk approximation). Taken together, the effects of rapid decay and disc thickness are not expected to make $Q_{\rm crit}$ deviate significantly from unity. 

\citet{rba:10} have argued that the value of $Q$ can depend on physical scale and in particular be lower than unity on a small scale even if $Q>1$ on a large scale. They assumed the scaling relations of $\Sigma_{\rm gas}\propto l^a$ and $\sigma_{\rm gas}\propto l^b$, where $l$ is physical scale over which $\Sigma_{\rm gas}$ and $\sigma_{\rm gas}$ are measured. They pointed out that if $b>(a+1)/2$ and $-2<a<1$, the disc is always unstable at small scales as $l\rightarrow0$. In this regime, the stability of the disc is no longer controlled by $Q$ values (see also \citealt{hr:12} for discs consisting of gas and stars, and \citealt{arg:15} for discussion in the context of simulations). Although we applied the smoothing of FWHM$=1.2~{\rm kpc}$ in order to focus on giant clumps of $\sim10^8$--$10^9~\Msun$, clump formation may start with Toomre instability on a smaller scale with a lower mass. Then, the seeds of clumps could rapidly accrete ambient mass and/or other clump seeds and may grow massive. If the $Q$ values are measured on a too-large scale, the instability for the seed clumps may be missed. Our results indicate that the proto-clump regions in our simulations can have $Q_{\rm 2comp}\gsim2$ at least on the physical scale of $\sim1~{\rm kpc}$. Fig. \ref{HalfSmoothing} in Appendix \ref{SmoothingDependence} indicate the same results for $Q_{\rm 2comp}$ when a smaller smoothing scale of FWHM$=0.59~{\rm kpc}$ is applied. In the Figures, although we see a number of small-scale regions where $Q_{\rm 2comp}<1$, the typical value of $Q_{\rm 2comp}$ in the inter-clump regions is still larger than $1.8$.
 
The Toomre analysis does not take into account radial migration of a proto-clump. If a gas cloud is on an eccentric orbit and has $Q>1$ at the peri-centre, the cloud may become unstable with $Q<1$ at the apo-centre since $\kappa$ generally decreases with a galactocentric distance while $\sigma$ and $\Sigma$ would remain constant within the proto-clump. At least in the cases of Fig. \ref{ProtoClumpCase1} and \ref{ProtoClumpCase2}, however, the proto-clumps do not seem to be on highly eccentric orbits.

\subsection{Clump formation by other instabilities}
\label{Possibility3}
Clump formation could be governed by other mechanisms of instability. Although a detailed study of such mechanisms is beyond the scope of this paper, we discuss various possibilities.

\subsubsection{Stimulus-induced formation}
\label{Stimuli}
It is important to note that the high-redshift galaxies are subject to a sequence of external perturbations and accretion, e.g. in the form of gas inflows and (mostly minor) mergers. The inflowing gas streams can have slow or retro-grade rotation velocities with respect to the disc rotation and generate patches of low $\overline{v_{\phi}}$ (i.e. low $\kappa$), where the centrifugal force in rotating frame may fail to support the massive proto-clumps against gravitational collapse.

At high redshift, the time between subsequent mergers can be shorter than the disc orbital time (the ratio of these times being $\propto (1+z)^{-1}$), therefore the disc is continuously perturbed by the mergers. To testify, in our simulated high-redshift clumpy discs, the fraction of ex-situ clumps that join the disc as minor mergers is found to be comparable to the fraction of in-situ clumps formed by disc instability \citep[][2015, in preparation]{mdc:13}. These external perturbations may stimulate the disc instability and be responsible for maintaining the disc in the non-linearly perturbed state with $Q$ significantly above unity.

One of the possibilities is that the clump formation with a high $Q$ is due to excessive compressive modes in the gas-disc turbulence. In unperturbed equilibrium, the compressive modes (of non-zero divergence) are expected to carry half the turbulent kinetic energy that is carried by the solenoidal modes (of zero divergence). If, at least locally, the compressive modes contribute a larger fraction of the overall velocity-dispersion field of gas, the $\sigma_{\rm gas}$ that enters $Q_{\rm gas}$ would be high, making $Q_{\rm 2comp}$ higher than unity, but the compression would help induce local collapse into clumps, instead of providing pressure support against the collapse. Preliminary tests indicate that the compressive modes are indeed excessive in our simulated galaxies (Mandelker et al. 2015, in preparation). The origin of compressive modes of turbulence could either be external or internal. Tidal effects between galaxies could be compressive in various circumstances, e.g. in galaxy mergers, and thus induce the turbulence compressive modes \citep{ddh:03,j:13,rbk:14}. Alternatively, pre-existing clumps in a disc may induce compression in a wake behind them.

In the standard two-dimensional Toomre analysis, the disc is assumed to be relatively thin in the sense that the predicted scale of Toomre instability is comparable to the disc scale height determined by the vertical velocity dispersion. In `discs' where the proto-clumps are significantly smaller than the disc scale height, the nature of the instability is modified to a three-dimensional instability. \citet{eh:15} have argued that the star formation in dwarf irregular galaxies in the local Universe may be governed by such a three-dimensional instability state. In our simulated high-redshift galaxies, the sizes of the giant proto-clumps are comparable to the disc thickness, but smaller proto-clumps may be smaller than the disc scale height and thus possibly subject to the three-dimensional instability.

In the discussion below, we utilize our simulations for preliminary tests of additional possibilities for instability mechanisms, to be studied in more detail in forthcoming papers.

\subsubsection{Swing amplification}
\label{SwingAmplification}
Swing amplification is a mechanism that enhances the density contrast of a spiral structure by transition from a leading to a trailing arm \citep{t:81}. Although well-organized spiral arms are not common features in our high-redshift simulations, they do occasionally show features like spiral arms that similar substructures may cause swing amplification. It has been known that the efficiency of swing amplification is characterized by an $X$ parameter,
\begin{equation}
  X\equiv\frac{2\pi R}{m\lambda_{\rm crit}}=\frac{\kappa^2R}{2\pi G\Sigma m},
  \label{X-parameter}
\end{equation} 
where $m$ is the number of modes, and $\lambda_{\rm crit}=4\pi^2 G\Sigma/\kappa^2$.\footnote{$\lambda_{\rm crit}$ corresponds to the longest unstable wavelength in a zero-pressure disc. In addition, a single-component disc with $Q=1$ becomes the most unstable at a wavelength of $\simeq\lambda_{\rm crit}/2$ \citep{bt:08}.} The $X$ parameter corresponds to the cotangent of the pitch angle for waves propagating radially with the critical wavenumber $k_{\rm crit}\equiv2\pi/\lambda_{\rm crit}$. Swing amplification becomes most efficient when $1\lsim X\lsim3$ \citep{t:81}. Here we adopt $X=2$, where swing amplification is expected to gain the maximum density enhancement for the mode 
\begin{equation}
  m=\frac{\kappa^2R}{4\pi G\Sigma}.
  \label{mode}
\end{equation} 

\begin{figure}
  \includegraphics[width=\hsize]{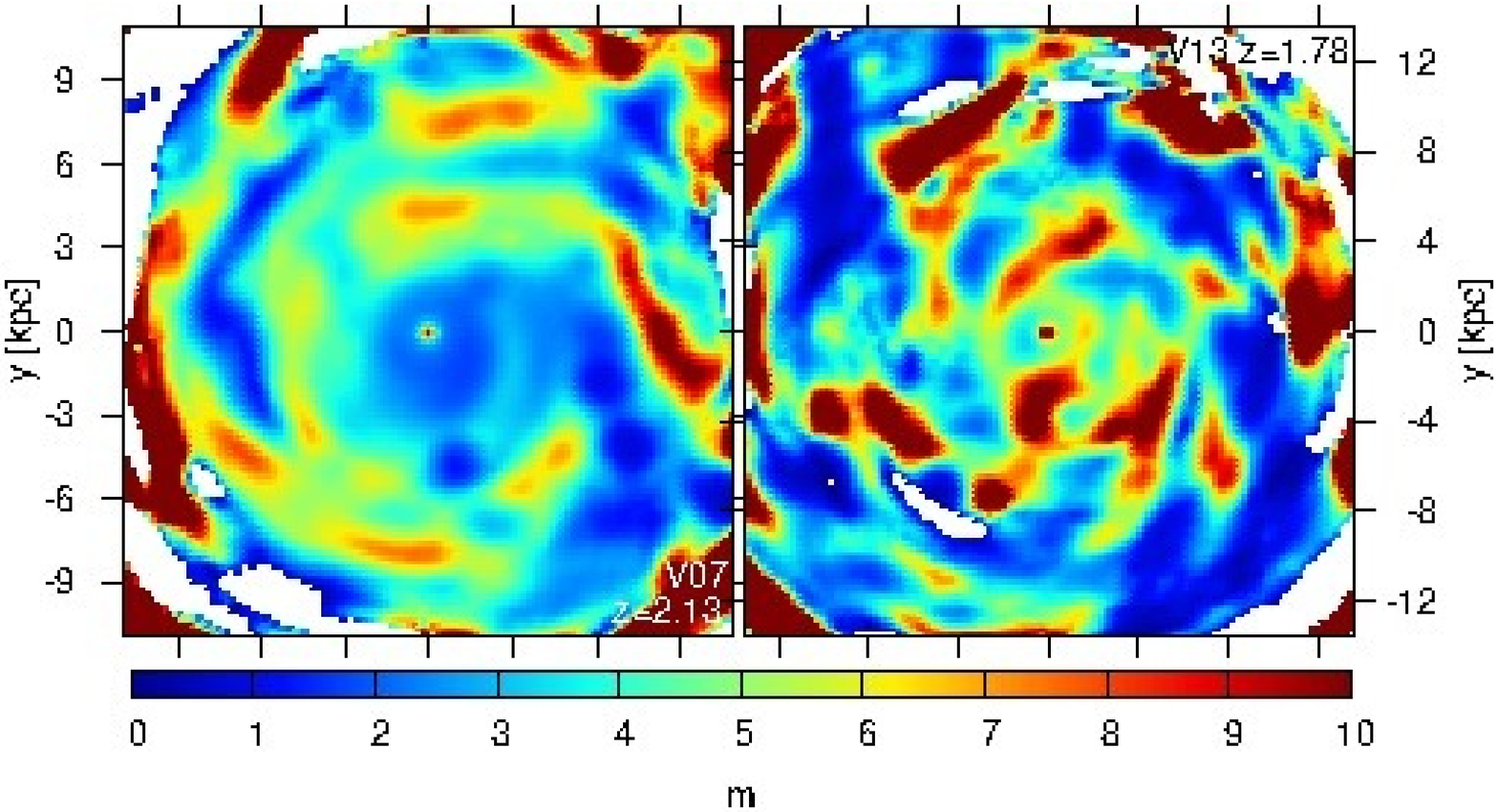}
  \includegraphics[width=\hsize]{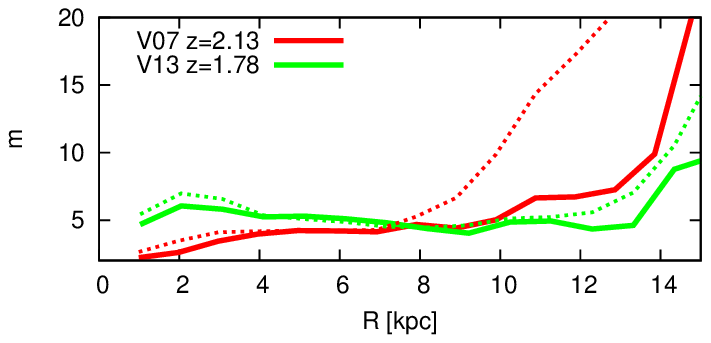}
  \caption{The top panels show maps of $m$ in Eq. (\ref{mode}) in the snapshots of V07 at $z=2.13$ (left) and V13 at $z=1.78$ (right). In these calculations, $\Sigma$ and $\kappa$ are redefined as $\Sigma\equiv\Sigma_{\rm Gas}+\Sigma_{\rm Star}$ and $\kappa\equiv(\kappa_{\rm Gas}\Sigma_{\rm Gas}+\kappa_{\rm Star}\Sigma_{\rm Star})/\Sigma$ on each position. The surface density maps of the snapshots are shown in Fig. \ref{QMaps_V07_z2.12} (V07 at $z=2.13$) and Fig. \ref{QMaps_V13_z1.78} (V13 at $z=1.78$). The bottom panel indicates radial profiles of the annularly averaged $m(R)$. The thick solid lines are the results using $\overline{v_{\phi}}$ to measure $\kappa_{\rm Gas}$ and $\kappa_{\rm Star}$ separately (see \S\ref{howtokappa}). The thin dotted-lines are the results using $v_{\rm circ}\equiv\sqrt{GM(<r)/r}$ to measure $\kappa$.}
  \label{ToomreX}
\end{figure}
The top panels of Fig. \ref{ToomreX} show maps of $m$ based on Eq. (\ref{mode}) in the snapshots of V07 at $z=2.13$ and V13 at $z=1.78$. We see that the clumpy regions have low values, $m\lsim3$. This implies that, if clumps form in amplified spiral-arm-like structures, they prefer a small number of such structures swinging in the disc. The bottom panel of Fig. \ref{ToomreX} shows the radial profiles of $m$, which seem similar among the two galaxies. We see $m\simeq5$ at most of radii, with $m$ slightly increasing with radius in V07. For the other galaxies listed in Table \ref{CosList}, we confirmed that $m=2$--$10$ in the disc regions during their clumpy phases. This confirms the impression of a preference for a fairly low $m$ for clump formation by swing amplification.

On the other hand, if a clump forms in a spiral arm at a radius $R$, the azimuthal wavelength corresponding to a proto-clump would be $\lambda_{\rm pc}\sim2\pi R/m$. If we expect $\lambda_{\rm cl}=4l_{\rm pc}\sim2~{\rm kpc}$ as in \S\ref{smoothing}, and if typical clump formation cites are at $R\sim8{\rm kpc}$, the number of modes would be estimated to be $m\sim25$. This estimation predicts $m$ significantly greater than those calculated from Eq. (\ref{mode}) in our simulations, shown in Fig. \ref{ToomreX}. In the bottom panel of Fig. \ref{ToomreX}, V07 has $m\gsim20$ increasing steeply at $R\sim15~{\rm kpc}$ ($R/R_{\rm vir}\simeq0.15$) since the surface density $\Sigma$ is truncated at the edge of the disc. However, Fig. \ref{Rhist} shows that the formation of the clumps mainly occurs at the inner radii $R/R_{\rm vir}<0.15$. Hence, the formation of giant clumps at the inner region cannot be attributed to the swing amplification expected from the abrupt increase of $m$ in the outer edge. We confirmed that although such steep increase of $m$ due to the truncation of $\Sigma$ at outer edges can be seen in all snapshots of our simulations, the formation of clumps generally occurs in inner regions where $m$ is low.

This indicates that clump formation could not be driven by swing amplification in our simulations. Another reason for disfavoring this scenario is our finding that $Q\gsim2$ in clump-forming regions of the disc in our cosmological simulations, while it has been proposed that the swing amplification becomes inefficient when $Q\gsim2$ \citep{t:81}. 

\subsubsection{Instability for non-axisymmetric perturbations}
\label{non-axi}
Even if a disc is stable for axisymmetric perturbations, it may be unstable for non-axisymmetric perturbations. \citet{lh:78} have performed a linear perturbation analysis of non-axisymmetric disturbance in a cold disc and derived a necessary condition for the instability in terms of the potential vorticity $f$ having at least one maximum or minimum as a function of radius $R$ \citep[see also][]{lc:99,lh:13}. The potential vorticity $f$ is defined to be
\begin{equation}
  f\equiv\frac{\Sigma\Omega}{\kappa^2},
  \label{Vortensity}
\end{equation} 
where $\Omega$ is angular velocity in a disc. The potential vorticity $f$ is positive, and it asymptotically approaches $\simeq0$ at a sufficiently large radius where $\Sigma\simeq0$. Therefore, it can be expected, in other words, that a disc would be stable for non-axisymmetric perturbation if $f(R)$ is constant or decreases monotonically with radius. Studies using numerical simulations have indicated that this instability can form spiral arms in $N$-body disc galaxies \citep{sk:91} and can grow into local clumpy structures with high densities in Keplarian gas discs \citep[e.g.][]{mcv:10,mkc:12}. One may expect that the non-axisymmetric instability could trigger clump formation in our simulations, based on the analysis of \citet{lh:13} which argued that such instability can occur even if $Q>1$.

\begin{figure}
  \includegraphics[width=\hsize]{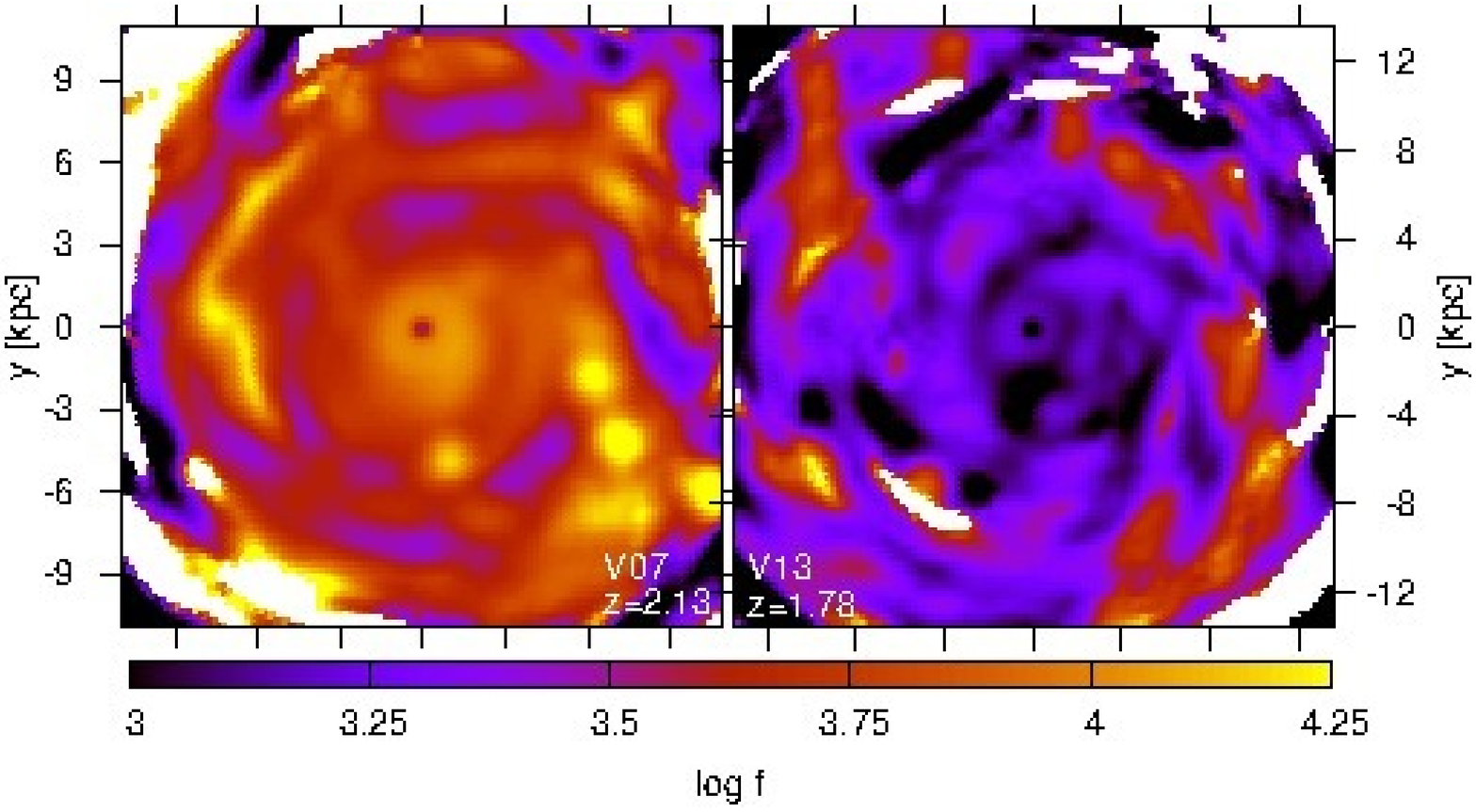}
  \includegraphics[width=\hsize]{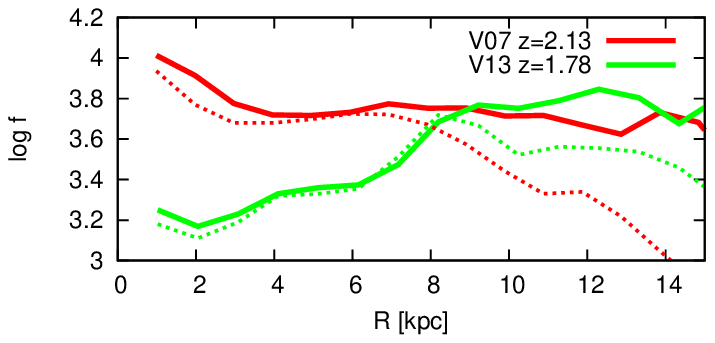}
  \caption{The top panels show maps of $f$ based on Eq. (\ref{Vortensity}) in V07 at $z=2.13$ (left) and V13 at $z=1.78$ (right). The values of $\Sigma$, $\kappa$ and $\Omega$ are redefined as $\Sigma\equiv\Sigma_{\rm Gas}+\Sigma_{\rm Star}$ and the density-weighted averaged values of the Gas and Star components for $\kappa$ and $\Omega$ on each position. The surface density maps of the snapshots are shown in Fig. \ref{QMaps_V07_z2.12} (V07 at $z=2.13$) and Fig. \ref{QMaps_V13_z1.78} (V13 at $z=1.78$). The bottom panel indicates radial profiles of the annularly averaged $f(R)$. The thick solid lines are the results using $\overline{v_{\phi}}$ to measure $\kappa_{\rm Gas}$ and $\kappa_{\rm Star}$ separately (see \S\ref{kappa}). The thin dotted-lines are the results using $v_{\rm circ}\equiv\sqrt{GM(<r)/r}$ to measure $\kappa$. $\Omega\equiv\overline{v_{\phi}}/R$ in both cases. The units of $f$ are arbitrary.}
  \label{VortensityMap}
\end{figure}
The top panels of Fig. \ref{VortensityMap} show maps of the potential vorticity $f$ based on Eq. (\ref{Vortensity}) in the snapshots of V07 at $z=2.13$ and V13 at $z=1.78$. The normalization of f is arbitrary since only the shape of the function is relevant. In V07 (top left), we notice high values of $f$ in the positions of clumps and at the galaxy centre. The peaks of $f$ values near the clumps appear to be local features, while the inter-clump regions seem to have $f$ profiles that are decreasing or nearly flat with radius. On the other hand, in V13 (top right), the galactic centre has a low value of $f$. Although the clumps tend to represent local peaks of $f$, as in V07, the regions with high $f$ are preferentially located at large radii, $R\sim8$--$10~{\rm kpc}$. The bottom panel of Fig. \ref{Vortensity} shows the radial profiles of $f(R)$ in the two snapshots. In V07, $f(R)$ is almost constant or slightly decreasing with radius. On the other hand, in V13, $f(R)$ is increasing with radius in $R\lsim12~{\rm kpc}$. Therefore, from the annularly averaged profiles of $f(R)$, V07 seems to be stable for non-axisymmetric perturbation, whereas V13 may be unstable. However, it should be recalled that the presence of minima or maxima in $f(R)$ is only a necessary condition for the instability. Hence, we cannot affirmatively conclude that even V13 is unstable for non-axisymmetric perturbations.

Besides the above analysis based on the potential vorticity $f$ proposed by \citet{lh:78}, a number of studies have addressed instabilities for non-axisymmetric perturbations \citep[for recent discussion, see \S5.2 of][and references therein]{rf:15,gg:12}. Using a linear perturbation theory relaxing the tight-winding approximation, i.e. $m\neq0$, \citet{gg:12} analysed gravitational instability in their disc model with a finite thickness. They found that non-axisymmetric $m\neq0$ perturbations are more unstable than axisymmetric $m=0$ ones \citep[see also][]{j:92}, and estimated the stability criterion to be $Q_{\rm crit}\simeq2$ for the non-axisymmetric perturbations. However, our results shown in \S\ref{ThinSlices} indicate that significant fractions of clumps start forming with $Q\gsim2$.

\subsection{On the observational determinations of $Q$}
\label{Observations}

As we mentioned in \S\ref{Intro}, \citet{gnj:11,gfl:14} produced two-dimensional maps of $Q$ for their observed massive clumpy discs at $z\sim2$. They estimated $Q$ for the gas using kinematic modeling based on H$\alpha$ spectroscopic measurements, and estimating gas surface densities based on the SFR deduced from H$\alpha$ intensity via the Kennicutt-Schmidt law \citep{s:59,k:89}. They found high $Q$ values in the central regions dominated by bulges, and values of $Q$ typically below unity in the extended discs and ring-like structures, which host giant clumps. 

  
In \S\ref{Qmapsingalaxies}, our analysis of the simulated galaxies typically reveals higher values of $Q$ in the extended, clumpy discs, when smoothed on a similar scale as the observed discs. The difference is pronounced mostly in simulated galaxies where the gas fraction is lower than the estimates for the mean gas fractions in the observed galaxies at $z\sim2$. For example, we have $f_{\rm gas}\sim0.2$ in V07 and V19 (see Table \ref{CosList}), whereas $f_{\rm gas}$ ranges from $0.2$ to $0.6$ in observed star-forming galaxies at this redshift range \citep{t:10,dbw:10,tkt:13II}. In our simulated galaxies where the gas fraction is higher (e.g. $f_{\rm gas}\sim0.4$ in V08 and V13), the global values of $Q$ are more compatible with the observed estimates.  


We should be aware of the fact that we clearly face a fundamental problem when trying to estimate $Q$ either in observed or simulated clumpy discs and attempting to interpret the result in terms of Toomre instability. The parameter $Q$ conceptually refers to linear instability of small-amplitude perturbations in a uniformly rotating background disc. On the other hand, the clumpy discs that we inspect are already non-linearly perturbed with the overdensities in the clumps exceeding a few tens, and the underdensities between the clumps are large accordingly. Since $\kappa$ as well as $\sigma$ are not very different between the compact clumps and outside them (Fig. \ref{Ingredients_V07_z2.12}, and \citealt[][]{cdm:11}), $Q$ is driven to low values predominantly by the non-linearly high $\Sigma$ within the clumps, independently of the detailed origin of the clumps. This limits our ability to address the Toomre-instability origin of the clumps directly from the observations or the simulations at given snapshots. However, the great virtue of the simulations is that they enable us to trace each clump back to its phase as a small-amplitude perturbation in the proto-clump Lagrangian patch. Our most meaningful result from the simulations is that $Q$ is larger than $2$--$3$ in some of these proto-clump regions. Such an analysis is infeasible in the observed galaxies.


It is interesting to note that the estimates of $Q$ in the solar neighbourhood within the Milky way range from $1$ to $3$ \citep{r:01,bt:08}. Other local spiral galaxies also indicate $Q\sim2$ \citep[e.g.][]{rf:13,wab:14}. These values are similar to the high values of $Q$ measured in our simulated high-redshift galaxies. However, the high-redshift clumpy discs are different from the local discs. The gas fraction at high redshift is higher (typically $f_{\rm gas}\gsim0.2$ versus $f_{\rm gas}<0.1$ locally), with fresh gas accreting continuously. The high-redshift disc perturbations are dominated by significantly more massive clumps and large-scale structures that largely follow the inflowing streams. In contrast, the low-redshift discs show much less massive clumps (Giant Molecular Clouds) that are embedded in well-organized spiral arms, and they frequently show galactic bars. These differences indicate that the high-redshift disc instability is associated with more violent dynamical processes as opposed to the secular processes in the local galaxies.

\section{Conclusions and summary}
\label{Conclusions}
We performed a Toomre analysis of high-redshift clumpy galaxies in cosmological simulations. The local Toomre $Q$ parameter has been computed considering the two components of gas (plus young stars) and stars, smoothed with a Gaussian kernel of an FWHM$=1.2~{\rm kpc}$ to focus on giant clumps of $\sim10^8$--$10^9~{\rm M_\odot}$.  

The perturbations in the discs are highly non-linear over long periods, with compact giant clumps and extended structures, so they are clearly not in the idealized conditions of linear fluctuations in an isolated, uniform disc, where linear Toomre instability is supposed to be strictly valid. This makes the interpretation of $Q$ rather nontrivial, both in simulations and in observations. In particular, the values of $Q$ are measured to be smaller than unity in the clumps since the surface densities in the collapsed clumps are considerably higher than the inter-clump regions. This is, however, independent of whether Toomre instability plays a role in the formation of the clumps. Our finding is that generally $Q\gsim2$--$3$ in the inter-clump regions, where new clumps are apparently forming. More specifically, $Q$ is high in the proto-clump regions, which are about to give birth to new clumps. A correction for the large disc thickness would bring $Q$ to even higher values, further away from the critical value of unity that is supposed to characterize the instability. The high $Q$ values indicates that clump formation cannot be simply understood in terms of the standard Toomre instability with $Q_{\rm crit}\sim1$. It is likely that the Toomre $Q$ parameter is still an approximate indicator for instability, but one should understand why $Q_{\rm crit}$ is significantly larger than unity.

We consider possible scenarios for clump formation with the high $Q$ in the clumpy galaxies at high redshifts. Within the realm of Toomre instability, one possibility could be that the state with a high $Q$ value can be unstable due to rapid decay of the turbulence pressure on a timescale shorter than the internal crossing time of the proto-clump region \citep{e:11}. In addition, the values of $Q$ can depend on physical scale. Even if $Q>1$ on the large scale $\sim1~{\rm kpc}$, the disc can be unstable with $Q<1$ on a smaller scale \citep[e.g.][]{rba:10}. These mechanisms could in principle allow low-mass clumps to form, and then possibly grow to massive clumps via further accretion and mergers with other low-mass clumps. Another scenario is that the instability with high $Q$ is associated with non-axisymmetric perturbations \citep[e.g.][]{lh:78}. However, we have found no evidence in our simulations for any of the above yet.

Alternatively, the instability may be stimulated by external drivers, such as minor mergers, counter-rotating streams and a triaxial potential well. In particular, interactions between galaxies may generate tidal compression that would result in excessive compressive modes of turbulence, inducing the collapse of gas into clumps while contributing to a high $Q_{\rm gas}$ through a high overall velocity dispersion $\sigma_{\rm gas}$. We also note that such a compressive mode may be generated internally in the disc, with the pre-existing clumps inducing new clump formation in wakes behind them.

Clump formation under the conditions of $Q\gsim2$--$3$, as found in our cosmological simulations of high-redshift galaxies, poses an open challenge to theory. The VDI at high redshifts seems to be different from the secular instability in low-redshift discs, and it requires a new theoretical understanding. The possible mechanisms mentioned above are currently under investigation.  

\section*{Acknowledgments}

We thank the reviewer, Alessandro B. Romeo, for his fascinating suggestions and are grateful for stimulating discussions with Bruce G. Elmegreen, Mark Krumholz, Yusuke Tsukamoto and Sharon Lapiner. The simulations presented in this paper were performed at the National Energy Research Scientific Computing Center (NERSC), Lawrence Berkeley National Laboratory, and at NASA Advanced Supercomputing (NAS) at NASA Ames Reserach Center. Development and most of the analysis have been performed in the astro cluster at the Hebrew University of Jerusalem. This study was supported by ISF grant 24/12, by the I-CORE Program of the PBC, by ISF grant 1829/12, by BSF grant 2014-273, by NSF grants AST-1010033 and AST-1405962, and by a grant AYA2012-32295, and by a grant from the Hayakawa Sachio Fund awarded by the Astronomical Society of Japan. SI acknowledges support from the Lady Davis Fellowship at the Hebrew University of Jerusalem.


\begin{thebibliography}{}

\bibitem[\protect\citeauthoryear{{Abraham} \& {van den Bergh}}{{Abraham} \&
  {van den Bergh}}{2001}]{ab:01}
{Abraham} R.~G.,  {van den Bergh} S.,  2001, Science, 293, 1273

\bibitem[\protect\citeauthoryear{{Agertz}, {Romeo} \& {Grisdale}}{{Agertz}
  et~al.}{2015}]{arg:15}
{Agertz} O.,  {Romeo} A.~B.,    {Grisdale} K.,  2015, \mnras, 449, 2156

\bibitem[\protect\citeauthoryear{{Aumer}, {White} \& {Naab}}{{Aumer}
  et~al.}{2014}]{awn:14}
{Aumer} M.,  {White} S.~D.~M.,    {Naab} T.,  2014, \mnras, 441, 3679

\bibitem[\protect\citeauthoryear{{Bassett} et~al.,}{{Bassett}
  et~al.}{2014}]{bgf:14}
{Bassett} R.,  et~al., 2014, \mnras, 442, 3206

\bibitem[\protect\citeauthoryear{{Behrendt}, {Burkert} \&
  {Schartmann}}{{Behrendt} et~al.}{2015}]{bbs:14}
{Behrendt} M.,  {Burkert} A.,    {Schartmann} M.,  2015, \mnras, 448, 1007

\bibitem[\protect\citeauthoryear{{Binney} \& {Tremaine}}{{Binney} \&
  {Tremaine}}{2008}]{bt:08}
{Binney} J.,  {Tremaine} S.,  2008, Galactic Dynamics Second Edition.
Princeton Univ. Press, Princeton

\bibitem[\protect\citeauthoryear{{Bournaud}, {Elmegreen} \&
  {Elmegreen}}{{Bournaud} et~al.}{2007}]{bee:07}
{Bournaud} F.,  {Elmegreen} B.~G.,    {Elmegreen} D.~M.,  2007, ApJ, 670, 237

\bibitem[\protect\citeauthoryear{{Bournaud}, {Elmegreen} \&
  {Martig}}{{Bournaud} et~al.}{2009}]{bem:09}
{Bournaud} F.,  {Elmegreen} B.~G.,    {Martig} M.,  2009, ApJL, 707, L1

\bibitem[\protect\citeauthoryear{{Bournaud} et~al.,}{{Bournaud}
  et~al.}{2014}]{bpr:14}
{Bournaud} F.,  et~al., 2014, \apj, 780, 57

\bibitem[\protect\citeauthoryear{{Cacciato}, {Dekel} \& {Genel}}{{Cacciato}
  et~al.}{2012}]{cdg:12}
{Cacciato} M.,  {Dekel} A.,    {Genel} S.,  2012, \mnras, 421, 818

\bibitem[\protect\citeauthoryear{{Ceverino}, {Dekel} \& {Bournaud}}{{Ceverino}
  et~al.}{2010}]{cdb:10}
{Ceverino} D.,  {Dekel} A.,    {Bournaud} F.,  2010, MNRAS, 404, 2151

\bibitem[\protect\citeauthoryear{{Ceverino}, {Dekel}, {Mandelker}, {Bournaud},
  {Burkert}, {Genzel} \& {Primack}}{{Ceverino} et~al.}{2012}]{cdm:11}
{Ceverino} D.,  {Dekel} A.,  {Mandelker} N.,  {Bournaud} F.,  {Burkert} A.,
  {Genzel} R.,    {Primack} J.,  2012, \mnras, 420, 3490

\bibitem[\protect\citeauthoryear{{Ceverino}, {Dekel}, {Tweed} \&
  {Primack}}{{Ceverino} et~al.}{2015}]{cdt:14}
{Ceverino} D.,  {Dekel} A.,  {Tweed} D.,    {Primack} J.,  2015, \mnras, 447,
  3291

\bibitem[\protect\citeauthoryear{{Ceverino} \& {Klypin}}{{Ceverino} \&
  {Klypin}}{2009}]{ck:09}
{Ceverino} D.,  {Klypin} A.,  2009, \apj, 695, 292

\bibitem[\protect\citeauthoryear{{Ceverino}, {Klypin}, {Klimek},
  {Trujillo-Gomez}, {Churchill}, {Primack} \& {Dekel}}{{Ceverino}
  et~al.}{2014}]{ckk:14}
{Ceverino} D.,  {Klypin} A.,  {Klimek} E.~S.,  {Trujillo-Gomez} S.,
  {Churchill} C.~W.,  {Primack} J.,    {Dekel} A.,  2014, \mnras, 442, 1545

\bibitem[\protect\citeauthoryear{{Chabrier}}{{Chabrier}}{2005}]{c:05}
{Chabrier} G.,  2005, in {Corbelli} E.,  {Palla} F.,   {Zinnecker} H.,  eds,
  The Initial Mass Function 50 Years Later Vol.~327 of Astrophysics and Space
  Science Library, {The Initial Mass Function: from Salpeter 1955 to 2005}.
p.~41

\bibitem[\protect\citeauthoryear{{Daddi} et~al.,}{{Daddi}
  et~al.}{2010}]{dbw:10}
{Daddi} E.,  et~al., 2010, \apj, 713, 686

\bibitem[\protect\citeauthoryear{{Dekel}, {Devor} \& {Hetzroni}}{{Dekel}
  et~al.}{2003}]{ddh:03}
{Dekel} A.,  {Devor} J.,    {Hetzroni} G.,  2003, \mnras, 341, 326

\bibitem[\protect\citeauthoryear{{Dekel}, {Sari} \& {Ceverino}}{{Dekel}
  et~al.}{2009}]{dsc:09}
{Dekel} A.,  {Sari} R.,    {Ceverino} D.,  2009, \apj, 703, 785

\bibitem[\protect\citeauthoryear{{Elmegreen}}{{Elmegreen}}{2011}]{e:11}
{Elmegreen} B.~G.,  2011, \apj, 737, 10

\bibitem[\protect\citeauthoryear{{Elmegreen}, {Bournaud} \&
  {Elmegreen}}{{Elmegreen} et~al.}{2008}]{ebe:08}
{Elmegreen} B.~G.,  {Bournaud} F.,    {Elmegreen} D.~M.,  2008, ApJ, 688, 67

\bibitem[\protect\citeauthoryear{{Elmegreen} \& {Elmegreen}}{{Elmegreen} \&
  {Elmegreen}}{2006}]{ee:06}
{Elmegreen} B.~G.,  {Elmegreen} D.~M.,  2006, \apj, 650, 644

\bibitem[\protect\citeauthoryear{{Elmegreen}, {Elmegreen}, {S{\'a}nchez
  Almeida}, {Mu{\~n}oz-Tu{\~n}{\'o}n}, {Dewberry}, {Putko}, {Teich} \&
  {Popinchalk}}{{Elmegreen} et~al.}{2013}]{ees:13}
{Elmegreen} B.~G.,  {Elmegreen} D.~M.,  {S{\'a}nchez Almeida} J.,
  {Mu{\~n}oz-Tu{\~n}{\'o}n} C.,  {Dewberry} J.,  {Putko} J.,  {Teich} Y.,
  {Popinchalk} M.,  2013, \apj, 774, 86

\bibitem[\protect\citeauthoryear{{Elmegreen}, {Elmegreen}, {Vollbach}, {Foster}
  \& {Ferguson}}{{Elmegreen} et~al.}{2005}]{eev:05}
{Elmegreen} B.~G.,  {Elmegreen} D.~M.,  {Vollbach} D.~R.,  {Foster} E.~R.,
  {Ferguson} T.~E.,  2005, \apj, 634, 101

\bibitem[\protect\citeauthoryear{{Elmegreen} \& {Hunter}}{{Elmegreen} \&
  {Hunter}}{2015}]{eh:15}
{Elmegreen} B.~G.,  {Hunter} D.~A.,  2015, \apj, 805, 145

\bibitem[\protect\citeauthoryear{{Elmegreen} \& {Struck}}{{Elmegreen} \&
  {Struck}}{2013}]{es:13}
{Elmegreen} B.~G.,  {Struck} C.,  2013, \apjl, 775, L35

\bibitem[\protect\citeauthoryear{{Elmegreen}, {Elmegreen} \&
  {Hirst}}{{Elmegreen} et~al.}{2004}]{eeh:04}
{Elmegreen} D.~M.,  {Elmegreen} B.~G.,    {Hirst} A.~C.,  2004, ApJL, 604, L21

\bibitem[\protect\citeauthoryear{{Feng}, {Lin}, {Wang} \& {Taam}}{{Feng}
  et~al.}{2014}]{flw:14}
{Feng} C.-C.,  {Lin} L.-H.,  {Wang} H.-H.,    {Taam} R.~E.,  2014, \apj, 785,
  103

\bibitem[\protect\citeauthoryear{{Feng}, {Di Matteo}, {Croft}, {Tenneti},
  {Bird}, {Battaglia} \& {Wilkins}}{{Feng} et~al.}{2015}]{fdc:15}
{Feng} Y.,  {Di Matteo} T.,  {Croft} R.,  {Tenneti} A.,  {Bird} S.,
  {Battaglia} N.,    {Wilkins} S.,  2015, \apjl, 808, L17

\bibitem[\protect\citeauthoryear{{Ferland}, {Korista}, {Verner}, {Ferguson},
  {Kingdon} \& {Verner}}{{Ferland} et~al.}{1998}]{fkv:98}
{Ferland} G.~J.,  {Korista} K.~T.,  {Verner} D.~A.,  {Ferguson} J.~W.,
  {Kingdon} J.~B.,    {Verner} E.~M.,  1998, \pasp, 110, 761

\bibitem[\protect\citeauthoryear{{Fisher} et~al.,}{{Fisher}
  et~al.}{2014}]{fgb:14}
{Fisher} D.~B.,  et~al., 2014, \apjl, 790, L30

\bibitem[\protect\citeauthoryear{{F{\"o}rster Schreiber} et~al.,}{{F{\"o}rster
  Schreiber}  et~al.}{2009}]{fgb:09}
{F{\"o}rster Schreiber} N.~M.,  et~al., 2009, \apj, 706, 1364

\bibitem[\protect\citeauthoryear{{Fujii}, {Baba}, {Saitoh}, {Makino}, {Kokubo}
  \& {Wada}}{{Fujii} et~al.}{2011}]{fbs:11}
{Fujii} M.~S.,  {Baba} J.,  {Saitoh} T.~R.,  {Makino} J.,  {Kokubo} E.,
  {Wada} K.,  2011, \apj, 730, 109

\bibitem[\protect\citeauthoryear{{Garland}, {Pisano}, {Mac Low}, {Kreckel},
  {Rabidoux} \& {Guzm{\'a}n}}{{Garland} et~al.}{2015}]{gpm:15}
{Garland} C.~A.,  {Pisano} D.~J.,  {Mac Low} M.-M.,  {Kreckel} K.,  {Rabidoux}
  K.,    {Guzm{\'a}n} R.,  2015, \apj, 807, 134

\bibitem[\protect\citeauthoryear{{Genzel} et~al.,}{{Genzel}
  et~al.}{2006}]{g:06}
{Genzel} R.,  et~al., 2006, \nat, 442, 786

\bibitem[\protect\citeauthoryear{{Genzel} et~al.,}{{Genzel}
  et~al.}{2008}]{g:08}
{Genzel} R.,  et~al., 2008, \apj, 687, 59

\bibitem[\protect\citeauthoryear{{Genzel} et~al.,}{{Genzel}
  et~al.}{2011}]{gnj:11}
{Genzel} R.,  et~al., 2011, ApJ, 733, 101

\bibitem[\protect\citeauthoryear{{Genzel} et~al.,}{{Genzel}
  et~al.}{2014}]{gfl:14}
{Genzel} R.,  et~al., 2014, \apj, 785, 75

\bibitem[\protect\citeauthoryear{{Goldreich} \& {Lynden-Bell}}{{Goldreich} \&
  {Lynden-Bell}}{1965}]{gl:65}
{Goldreich} P.,  {Lynden-Bell} D.,  1965, \mnras, 130, 97

\bibitem[\protect\citeauthoryear{{Governato}, {Willman}, {Mayer}, {Brooks},
  {Stinson}, {Valenzuela}, {Wadsley} \& {Quinn}}{{Governato}
  et~al.}{2007}]{gwm:07}
{Governato} F.,  {Willman} B.,  {Mayer} L.,  {Brooks} A.,  {Stinson} G.,
  {Valenzuela} O.,  {Wadsley} J.,    {Quinn} T.,  2007, \mnras, 374, 1479

\bibitem[\protect\citeauthoryear{{Griv} \& {Gedalin}}{{Griv} \&
  {Gedalin}}{2012}]{gg:12}
{Griv} E.,  {Gedalin} M.,  2012, \mnras, 422, 600

\bibitem[\protect\citeauthoryear{{Guo} et~al.,}{{Guo}  et~al.}{2015}]{gfb:14}
{Guo} Y.,  et~al., 2015, \apj, 800, 39

\bibitem[\protect\citeauthoryear{{Haardt} \& {Madau}}{{Haardt} \&
  {Madau}}{1996}]{hm:96}
{Haardt} F.,  {Madau} P.,  1996, \apj, 461, 20

\bibitem[\protect\citeauthoryear{{Hoffmann} \& {Romeo}}{{Hoffmann} \&
  {Romeo}}{2012}]{hr:12}
{Hoffmann} V.,  {Romeo} A.~B.,  2012, \mnras, 425, 1511

\bibitem[\protect\citeauthoryear{{Hu} \& {Sijacki}}{{Hu} \&
  {Sijacki}}{2015}]{hs:15}
{Hu} S.,  {Sijacki} D.,  2015, preprint (astro-ph/1507.01643)

\bibitem[\protect\citeauthoryear{{Inoue}}{{Inoue}}{2013}]{i:13}
{Inoue} S.,  2013, \aap, 550, A11

\bibitem[\protect\citeauthoryear{{Inoue} \& {Saitoh}}{{Inoue} \&
  {Saitoh}}{2011}]{is:11}
{Inoue} S.,  {Saitoh} T.~R.,  2011, \mnras, 418, 2527

\bibitem[\protect\citeauthoryear{{Inoue} \& {Saitoh}}{{Inoue} \&
  {Saitoh}}{2012}]{is:12}
{Inoue} S.,  {Saitoh} T.~R.,  2012, \mnras, 422, 1902

\bibitem[\protect\citeauthoryear{{Inoue} \& {Saitoh}}{{Inoue} \&
  {Saitoh}}{2014}]{is:14}
{Inoue} S.,  {Saitoh} T.~R.,  2014, \mnras, 441, 243

\bibitem[\protect\citeauthoryear{{Jog}}{{Jog}}{1992}]{j:92}
{Jog} C.~J.,  1992, \apj, 390, 378

\bibitem[\protect\citeauthoryear{{Jog}}{{Jog}}{1996}]{j:96}
{Jog} C.~J.,  1996, \mnras, 278, 209

\bibitem[\protect\citeauthoryear{{Jog}}{{Jog}}{2013}]{j:13}
{Jog} C.~J.,  2013, \mnras, 434, L56

\bibitem[\protect\citeauthoryear{{Jog} \& {Solomon}}{{Jog} \&
  {Solomon}}{1984a}]{js:84b}
{Jog} C.~J.,  {Solomon} P.~M.,  1984a, \apj, 276, 127

\bibitem[\protect\citeauthoryear{{Jog} \& {Solomon}}{{Jog} \&
  {Solomon}}{1984b}]{js:84}
{Jog} C.~J.,  {Solomon} P.~M.,  1984b, \apj, 276, 114

\bibitem[\protect\citeauthoryear{{Kennicutt} Jr.}{{Kennicutt}}{1989}]{k:89}
{Kennicutt} Jr. R.~C.,  1989, ApJ, 344, 685

\bibitem[\protect\citeauthoryear{{Komatsu} et~al.,}{{Komatsu}
  et~al.}{2009}]{komatsu:09}
{Komatsu} E.,  et~al., 2009, \apjs, 180, 330

\bibitem[\protect\citeauthoryear{{Kravtsov}}{{Kravtsov}}{2003}]{k:03}
{Kravtsov} A.~V.,  2003, \apjl, 590, L1

\bibitem[\protect\citeauthoryear{{Kravtsov}, {Klypin} \& {Khokhlov}}{{Kravtsov}
  et~al.}{1997}]{kkk:97}
{Kravtsov} A.~V.,  {Klypin} A.~A.,    {Khokhlov} A.~M.,  1997, \apjs, 111, 73

\bibitem[\protect\citeauthoryear{{Leroy}, {Walter}, {Brinks}, {Bigiel}, {de
  Blok}, {Madore} \& {Thornley}}{{Leroy} et~al.}{2008}]{lwb:08}
{Leroy} A.~K.,  {Walter} F.,  {Brinks} E.,  {Bigiel} F.,  {de Blok} W.~J.~G.,
  {Madore} B.,    {Thornley} M.~D.,  2008, \aj, 136, 2782

\bibitem[\protect\citeauthoryear{{Li}, {Mac Low} \& {Klessen}}{{Li}
  et~al.}{2005a}]{lmk:05a}
{Li} Y.,  {Mac Low} M.-M.,    {Klessen} R.~S.,  2005a, \apjl, 620, L19

\bibitem[\protect\citeauthoryear{{Li}, {Mac Low} \& {Klessen}}{{Li}
  et~al.}{2005b}]{lmk:05b}
{Li} Y.,  {Mac Low} M.-M.,    {Klessen} R.~S.,  2005b, \apj, 626, 823

\bibitem[\protect\citeauthoryear{{Li}, {Mac Low} \& {Klessen}}{{Li}
  et~al.}{2006}]{lmk:06}
{Li} Y.,  {Mac Low} M.-M.,    {Klessen} R.~S.,  2006, \apj, 639, 879

\bibitem[\protect\citeauthoryear{{Lovelace} \& {Hohlfeld}}{{Lovelace} \&
  {Hohlfeld}}{1978}]{lh:78}
{Lovelace} R.~V.~E.,  {Hohlfeld} R.~G.,  1978, \apj, 221, 51

\bibitem[\protect\citeauthoryear{{Lovelace} \& {Hohlfeld}}{{Lovelace} \&
  {Hohlfeld}}{2013}]{lh:13}
{Lovelace} R.~V.~E.,  {Hohlfeld} R.~G.,  2013, \mnras, 429, 529

\bibitem[\protect\citeauthoryear{{Lovelace}, {Li}, {Colgate} \&
  {Nelson}}{{Lovelace} et~al.}{1999}]{lc:99}
{Lovelace} R.~V.~E.,  {Li} H.,  {Colgate} S.~A.,    {Nelson} A.~F.,  1999,
  \apj, 513, 805

\bibitem[\protect\citeauthoryear{{Mandelker}, {Dekel}, {Ceverino}, {Tweed},
  {Moody} \& {Primack}}{{Mandelker} et~al.}{2014}]{mdc:13}
{Mandelker} N.,  {Dekel} A.,  {Ceverino} D.,  {Tweed} D.,  {Moody} C.~E.,
  {Primack} J.,  2014, \mnras, 443, 3675

\bibitem[\protect\citeauthoryear{{Martig}, {Bournaud}, {Teyssier} \&
  {Dekel}}{{Martig} et~al.}{2009}]{mbt:09}
{Martig} M.,  {Bournaud} F.,  {Teyssier} R.,    {Dekel} A.,  2009, \apj, 707,
  250

\bibitem[\protect\citeauthoryear{{Meheut}, {Casse}, {Varniere} \&
  {Tagger}}{{Meheut} et~al.}{2010}]{mcv:10}
{Meheut} H.,  {Casse} F.,  {Varniere} P.,    {Tagger} M.,  2010, \aap, 516, A31

\bibitem[\protect\citeauthoryear{{Meheut}, {Keppens}, {Casse} \&
  {Benz}}{{Meheut} et~al.}{2012}]{mkc:12}
{Meheut} H.,  {Keppens} R.,  {Casse} F.,    {Benz} W.,  2012, \aap, 542, A9

\bibitem[\protect\citeauthoryear{{Michikoshi} \& {Kokubo}}{{Michikoshi} \&
  {Kokubo}}{2014}]{mk:14}
{Michikoshi} S.,  {Kokubo} E.,  2014, \apj, 787, 174

\bibitem[\protect\citeauthoryear{{Moody}, {Guo}, {Mandelker}, {Ceverino},
  {Mozena}, {Koo}, {Dekel} \& {Primack}}{{Moody} et~al.}{2014}]{mgm:14}
{Moody} C.~E.,  {Guo} Y.,  {Mandelker} N.,  {Ceverino} D.,  {Mozena} M.,  {Koo}
  D.~C.,  {Dekel} A.,    {Primack} J.,  2014, \mnras, 444, 1389

\bibitem[\protect\citeauthoryear{{Morozov}}{{Morozov}}{1981}]{m:81}
{Morozov} A.~G.,  1981, Soviet Astronomy Letters, 7, 5

\bibitem[\protect\citeauthoryear{{Murata} et~al.,}{{Murata}
  et~al.}{2014}]{mkt:14}
{Murata} K.~L.,  et~al., 2014, \apj, 786, 15

\bibitem[\protect\citeauthoryear{{Noguchi}}{{Noguchi}}{1996}]{n:96}
{Noguchi} M.,  1996, ApJ, 469, 605

\bibitem[\protect\citeauthoryear{{Noguchi}}{{Noguchi}}{1998}]{n:98}
{Noguchi} M.,  1998, Nat, 392, 253

\bibitem[\protect\citeauthoryear{{Noguchi}}{{Noguchi}}{1999}]{n:99}
{Noguchi} M.,  1999, ApJ, 514, 77

\bibitem[\protect\citeauthoryear{{Nordstr{\"o}m} et~al.,}{{Nordstr{\"o}m}
  et~al.}{2004}]{n:04}
{Nordstr{\"o}m} B.,  et~al., 2004, \aap, 418, 989

\bibitem[\protect\citeauthoryear{{Obreschkow} et~al.,}{{Obreschkow}
  et~al.}{2015}]{ogb:15}
{Obreschkow} D.,  et~al., 2015, preprint (astro-ph/1508.04768)

\bibitem[\protect\citeauthoryear{{Perez}, {Valenzuela}, {Tissera} \&
  {Michel-Dansac}}{{Perez} et~al.}{2013}]{pvt:13}
{Perez} J.,  {Valenzuela} O.,  {Tissera} P.~B.,    {Michel-Dansac} L.,  2013,
  \mnras, 436, 259

\bibitem[\protect\citeauthoryear{{Puech}}{{Puech}}{2010}]{p:10}
{Puech} M.,  2010, \mnras, 406, 535

\bibitem[\protect\citeauthoryear{{Rafikov}}{{Rafikov}}{2001}]{r:01}
{Rafikov} R.~R.,  2001, \mnras, 323, 445

\bibitem[\protect\citeauthoryear{{Renaud}, {Bournaud}, {Kraljic} \&
  {Duc}}{{Renaud} et~al.}{2014}]{rbk:14}
{Renaud} F.,  {Bournaud} F.,  {Kraljic} K.,    {Duc} P.-A.,  2014, \mnras, 442,
  L33

\bibitem[\protect\citeauthoryear{{Romeo}}{{Romeo}}{1985}]{r:85}
{Romeo} A.~B.,  1985, Master's thesis, Tesi di Laurea, University of Pisa and
  Scuola Normale Superiore, Pisa, Italy

\bibitem[\protect\citeauthoryear{{Romeo}}{{Romeo}}{1992}]{r:92}
{Romeo} A.~B.,  1992, \mnras, 256, 307

\bibitem[\protect\citeauthoryear{{Romeo}}{{Romeo}}{1994}]{r:94}
{Romeo} A.~B.,  1994, \aap, 286, 799

\bibitem[\protect\citeauthoryear{{Romeo} \& {Agertz}}{{Romeo} \&
  {Agertz}}{2014}]{ra:14}
{Romeo} A.~B.,  {Agertz} O.,  2014, \mnras, 442, 1230

\bibitem[\protect\citeauthoryear{{Romeo}, {Burkert} \& {Agertz}}{{Romeo}
  et~al.}{2010}]{rba:10}
{Romeo} A.~B.,  {Burkert} A.,    {Agertz} O.,  2010, \mnras, 407, 1223

\bibitem[\protect\citeauthoryear{{Romeo} \& {Falstad}}{{Romeo} \&
  {Falstad}}{2013}]{rf:13}
{Romeo} A.~B.,  {Falstad} N.,  2013, \mnras, 433, 1389

\bibitem[\protect\citeauthoryear{{Romeo} \& {Fathi}}{{Romeo} \&
  {Fathi}}{2015}]{rf:15}
{Romeo} A.~B.,  {Fathi} K.,  2015, \mnras, 451, 3107

\bibitem[\protect\citeauthoryear{{Romeo} \& {Wiegert}}{{Romeo} \&
  {Wiegert}}{2011}]{rw:11}
{Romeo} A.~B.,  {Wiegert} J.,  2011, \mnras, 416, 1191

\bibitem[\protect\citeauthoryear{{Schmidt}}{{Schmidt}}{1959}]{s:59}
{Schmidt} M.,  1959, ApJ, 129, 243

\bibitem[\protect\citeauthoryear{{Sellwood} \& {Kahn}}{{Sellwood} \&
  {Kahn}}{1991}]{sk:91}
{Sellwood} J.~A.,  {Kahn} F.~D.,  1991, \mnras, 250, 278

\bibitem[\protect\citeauthoryear{{Shapiro}, {Genzel} \& {F{\"o}rster
  Schreiber}}{{Shapiro} et~al.}{2010}]{sgf:10}
{Shapiro} K.~L.,  {Genzel} R.,    {F{\"o}rster Schreiber} N.~M.,  2010, \mnras,
  403, L36

\bibitem[\protect\citeauthoryear{{Shlosman} \& {Noguchi}}{{Shlosman} \&
  {Noguchi}}{1993}]{SN:93}
{Shlosman} I.,  {Noguchi} M.,  1993, \apj, 414, 474

\bibitem[\protect\citeauthoryear{{Tacconi} et~al.,}{{Tacconi}
  et~al.}{2010}]{t:10}
{Tacconi} L.~J.,  et~al., 2010, \nat, 463, 781

\bibitem[\protect\citeauthoryear{{Tadaki}, {Kodama}, {Tanaka}, {Hayashi},
  {Koyama} \& {Shimakawa}}{{Tadaki} et~al.}{2014}]{tkt:13II}
{Tadaki} K.-i.,  {Kodama} T.,  {Tanaka} I.,  {Hayashi} M.,  {Koyama} Y.,
  {Shimakawa} R.,  2014, \apj, 780, 77

\bibitem[\protect\citeauthoryear{{Toomre}}{{Toomre}}{1964}]{t:64}
{Toomre} A.,  1964, ApJ, 139, 1217

\bibitem[\protect\citeauthoryear{{Toomre}}{{Toomre}}{1981}]{t:81}
{Toomre} A.,  1981, in {Fall} S.~M.,  {Lynden-Bell} D.,  eds, Structure and
  Evolution of Normal Galaxies {What amplifies the spirals}.
pp 111--136

\bibitem[\protect\citeauthoryear{{van den Bergh}, {Abraham}, {Ellis}, {Tanvir},
  {Santiago} \& {Glazebrook}}{{van den Bergh} et~al.}{1996}]{bae:96}
{van den Bergh} S.,  {Abraham} R.~G.,  {Ellis} R.~S.,  {Tanvir} N.~R.,
  {Santiago} B.~X.,    {Glazebrook} K.~G.,  1996, \aj, 112, 359

\bibitem[\protect\citeauthoryear{{Wang} \& {Silk}}{{Wang} \&
  {Silk}}{1994}]{ws:94}
{Wang} B.,  {Silk} J.,  1994, \apj, 427, 759

\bibitem[\protect\citeauthoryear{{Wang}, {Klessen}, {Dullemond}, {van den
  Bosch} \& {Fuchs}}{{Wang} et~al.}{2010}]{wkd:10}
{Wang} H.-H.,  {Klessen} R.~S.,  {Dullemond} C.~P.,  {van den Bosch} F.~C.,
  {Fuchs} B.,  2010, \mnras, 407, 705

\bibitem[\protect\citeauthoryear{{Weiner} et~al.,}{{Weiner}
  et~al.}{2006}]{w:06}
{Weiner} B.~J.,  et~al., 2006, \apj, 653, 1027

\bibitem[\protect\citeauthoryear{{Westfall}, {Andersen}, {Bershady},
  {Martinsson}, {Swaters} \& {Verheijen}}{{Westfall} et~al.}{2014}]{wab:14}
{Westfall} K.~B.,  {Andersen} D.~R.,  {Bershady} M.~A.,  {Martinsson} T.~P.~K.,
   {Swaters} R.~A.,    {Verheijen} M.~A.~W.,  2014, \apj, 785, 43

\bibitem[\protect\citeauthoryear{{Yang}, {Gruendl}, {Chu}, {Mac Low} \&
  {Fukui}}{{Yang} et~al.}{2007}]{ygc:07}
{Yang} C.-C.,  {Gruendl} R.~A.,  {Chu} Y.-H.,  {Mac Low} M.-M.,    {Fukui} Y.,
  2007, \apj, 671, 374

\bibitem[\protect\citeauthoryear{{Zolotov} et~al.,}{{Zolotov}
  et~al.}{2015}]{z:15}
{Zolotov} A.,  et~al., 2015, \mnras, 450, 2327

\end{thebibliography}

\appendix

\section[]{The robustness of the results}
\label{App}
In our Toomre analysis, we introduce several parameters that have to be set somewhat arbitrarily. Additionally, an alternative formulation for $Q_{\rm 2comp}$ has also been proposed. Here, we demonstrate the robustness of our results with respect to the parameters, the formulation and the settings of our analysis.

\subsection[]{Dependence on the formulation of $Q_{\rm 2comp}$}
\label{App1}
Stability conditions for axisymmetric perturbations in a multi-component disc have been discussed by a number of studies \citep[e.g.][]{m:81,r:85,ws:94,j:96,r:01}. \citet{js:84b,js:84} proposed a formulation of a two-component $Q$ parameter, in which their dispersion relation assumes a fluid-fluid disc system. \citet{rw:11} suggested an approximation of the formulation: Eq. (\ref{Q2comp}). Therefore, the two-component Q parameter we use in this study is based on the fluid-fluid dispersion relation proposed by \citet{js:84b,js:84}, in which the dispersion relation itself does not distinguish gas and stars, and this assumption is reflected as the symmetric form of $Q_{\rm 2comp}$ with respect to $Q_{\rm gas}$ and $Q_{\rm star}$ in Eq. (\ref{Q2comp}). 

Meanwhile, \citet{r:01} has also invented a different two-component model based on a star-fluid dispersion relation and using the same constant for $Q_{\rm gas}$ and $Q_{\rm star}$: $A_{\rm gas}=A_{\rm star}=\pi$. He discussed that his dispersion relation behaves different from that of \citet{js:84b,js:84} on short wavelengths at which stars can be more unstable than gas since stars lack gaseous pressure. $Q_{\rm 2comp}$ in his formulation is described for wavenumber $k$, using $\kappa_{\rm Gas}$ and $\kappa_{\rm Star}$ separately, as 
\begin{eqnarray}
  \frac{1}{Q_{\rm 2comp}} = \frac{2}{Q_{\rm Star}}\frac{1}{q_{\rm Star}}\left[1-\exp\left(-q_{\rm Star}^2\right)I_0\left(q_{\rm Star}^2\right)\right]&\nonumber\\
  + \frac{2}{Q_{\rm Gas}}\xi\frac{q_{\rm Gas}}{1+q_{\rm Gas}^2\xi^2},
\label{RafikovQ}
\end{eqnarray}
where $\xi\equiv\sigma_{\rm Gas}/\sigma_{\rm Star}$, $q_{\rm Gas}\equiv k\sigma_{\rm Star}/\kappa_{\rm Gas}$ and $q_{\rm Gas}\equiv k\sigma_{\rm Star}/\kappa_{\rm Star}$, and $I_0$ is the first-kind modified Bessel function of order zero. The minimum value of $Q_{\rm 2comp}$ is obtained at the wavenumber $k$ that gives the lowest value.
Although the original formulation of \citet{r:01} uses $\kappa$ measured from $v_{\rm circ}$, therefore $\kappa_{\rm Star}=\kappa_{\rm Gas}$, here we follow our main analysis method described in \S\ref{howtokappa} and compute $\kappa_{\rm Star}$ and $\kappa_{\rm Gas}$ from $\overline{v_{\phi}}$ for each component.

\begin{figure}
  \includegraphics[width=\hsize]{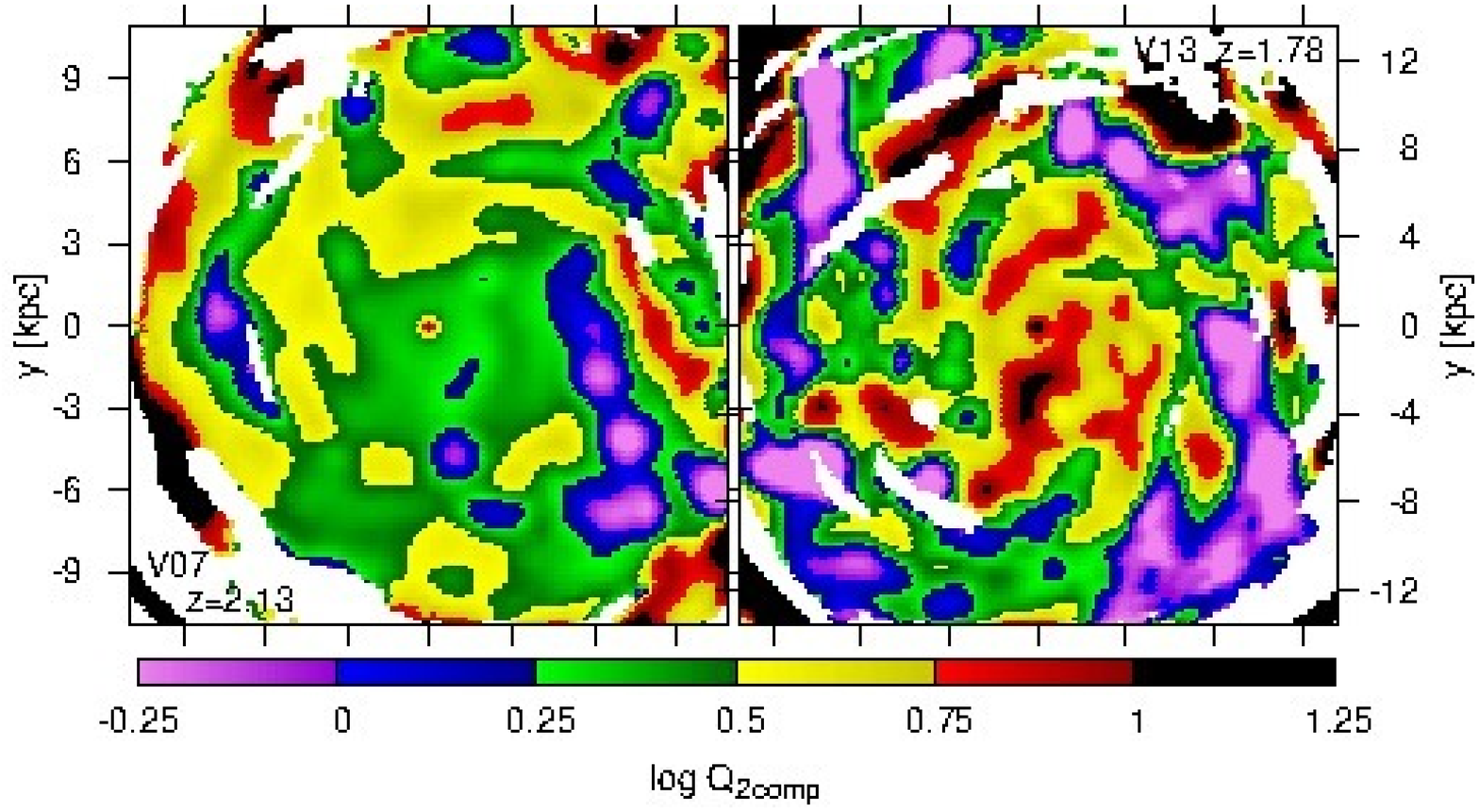}
  \caption{Maps of the minimum $Q_{\rm 2comp}$ using the formulation of \citet{r:01} in V07 at $z=2.13$ (left) and V13 at $z=1.78$ (right). The other settings of the analysis are the same as in \S\ref{Results}, and the thickness correction is not applied. The same figures but using the formulation of \citet[][Eq. \ref{Q2comp}]{rw:11} are seen in the bottom left panels of Fig. \ref{QMaps_V07_z2.12} and \ref{QMaps_V13_z1.78}, respectively.}
  \label{Rafikov}
\end{figure}
Fig. \ref{Rafikov} shows the lowest $Q_{\rm 2comp}$ given by Eq. (\ref{Rafikov}) in the snapshots of V07 at $z=2.13$ and V13 at $z=1.78$. In comparison with $Q_{\rm 2comp}$ using Eq. (\ref{Q2comp}) shown in the bottom left panels of Fig. \ref{QMaps_V07_z2.12} and \ref{QMaps_V13_z1.78}, they are almost indistinguishable. \citet{rf:13} also demonstrated that the differences in the $Q$ values obtained by the formulations of \citet{rw:11} and \citet{r:01} amount to less than ten per cent in the case of a two-component model using $\kappa$ measured from $v_{\rm circ}$. Therefore, we can conclude that our results are robust with respect to the formulation for computing $Q_{\rm 2comp}$.

\subsection[]{Dependence on how to measure $\kappa$}
\label{Spherical}
In this study, we obtain epicyclic frequency $\kappa$ using Eq. \ref{kappa} from local mean rotation velocity $\overline{v_{\phi}}$ for each of the Gas and Star components separately. As we mentioned in \S\ref{howtokappa}, $\kappa$ could be approximated by using circular velocity $v_{\rm circ}$ instead of $\overline{v_{\phi}}$. Measuring the circular velocities in cosmological simulations usually has to assume a spherical mass distribution as $v_{\rm circ}\equiv\sqrt{GM(<r)/r}$, where $M(<R)$ is the total mass enclosed within $r$. In this case, $\kappa_{\rm Gas}$ and $\kappa_{\rm Star}$ are unified.

\begin{figure}
  \includegraphics[width=\hsize]{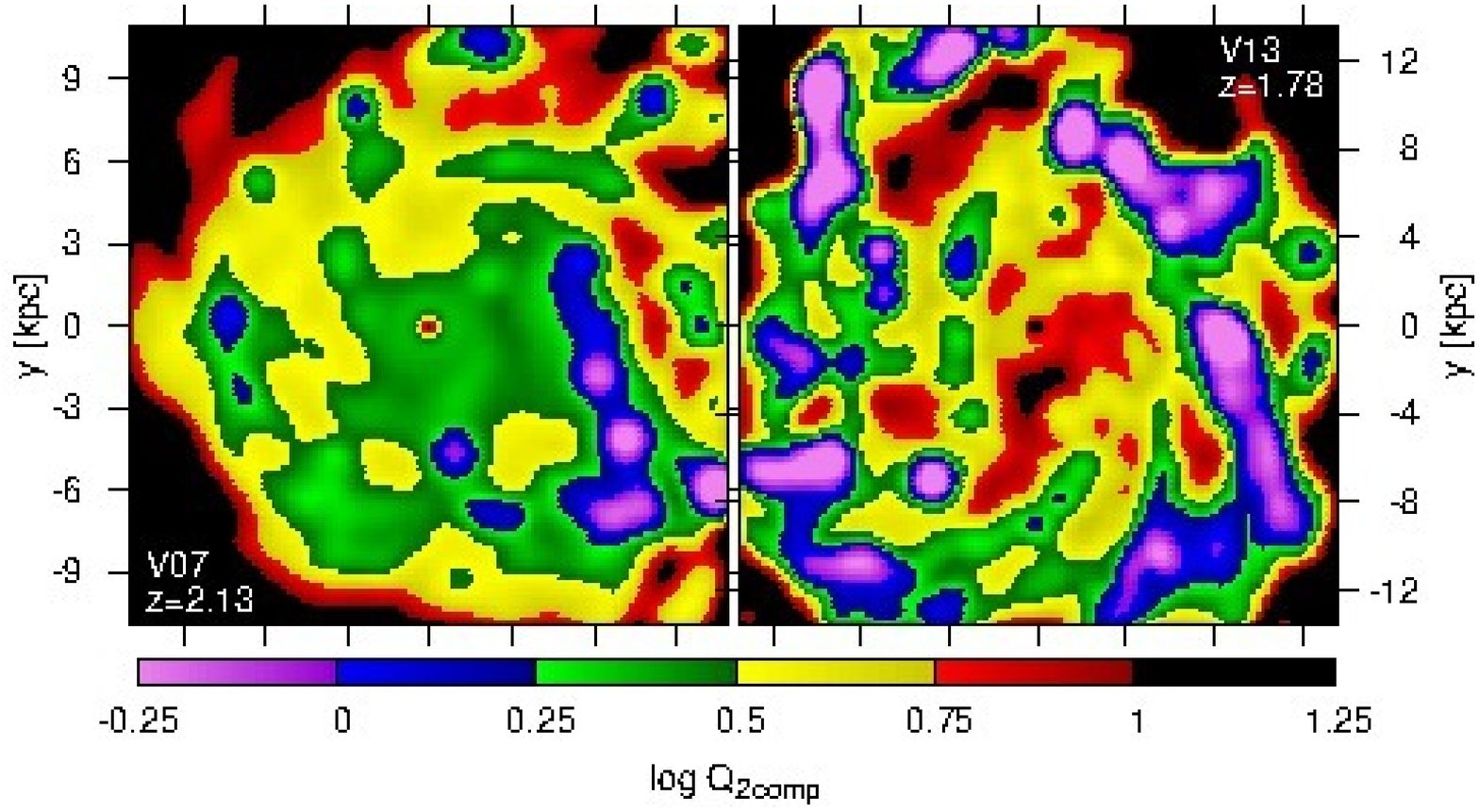}
  \caption{Maps of $Q_{\rm 2comp}$ using circular velocity, $v_{\rm circ}\equiv\sqrt{GM(<r)/r}$, for measuring $\kappa$ in V07 at $z=2.13$ (left) and V13 at $z=1.78$ (right). The same results but using $\overline{v_{\phi}}$ for measuring $\kappa$ are shown in the bottom left panels of Fig. \ref{QMaps_V07_z2.12} and \ref{QMaps_V13_z1.78}, respectively. The other parameters and formulation are the same as the results presented in \S\ref{Results}.}
  \label{Vcirc}
\end{figure}
Fig. \ref{Vcirc} shows the results of $Q_{\rm 2comp}$ using $v_{\rm circ}$ for measuring $\kappa$. As shown in the figure, $v_{\rm circ}$ does not cause $\kappa$ to be imaginary (see \S\ref{howtokappa}). In the comparison of the Figure with the left bottom panels in Fig. \ref{QMaps_V07_z2.12} and \ref{QMaps_V13_z1.78}, as expected, using $v_{\rm circ}$ generally leads $Q_{\rm 2comp}$ to higher values although only a few regions indicate slightly lower values. However, the systematic increase of $Q_{\rm 2comp}$ by using $v_{\rm circ}$ is not significant. Hence, it could be said that our results presented in \S\ref{Results} are robust with respect to the measurements of $\kappa$.

\subsection[]{Dependence on the bulge-star removal}
\label{Bulge}
Our analysis excludes stellar particles if they are classified into bulge stars according to the condition of $J_z/J_{\rm c}<0.7$ (see \S\ref{bulge}). The galaxies in our sample have stellar $B/T$ ratios between $0.3$ and $0.5$ (see Table \ref{CosList}), therefore their stellar masses are significantly reduced by the bulge-star removal applied in our analysis, which may affect the values of $Q_{\rm Star}$. The bulge stars generally have lower rotation velocities and higher radial velocity dispersions than the disc stars. Hence, by removing the bulge stars, $\Sigma_{\rm Star}$ decreases and $\kappa_{\rm Star}$ increases, therefore $Q_{\rm Star}$ may be expected to increase. On the other hand, since $\sigma_{\rm Star}$ decreases by the bulge-star removal, it is also expected that $Q_{\rm Star}$ may decrease.

\begin{figure}
  \includegraphics[width=\hsize]{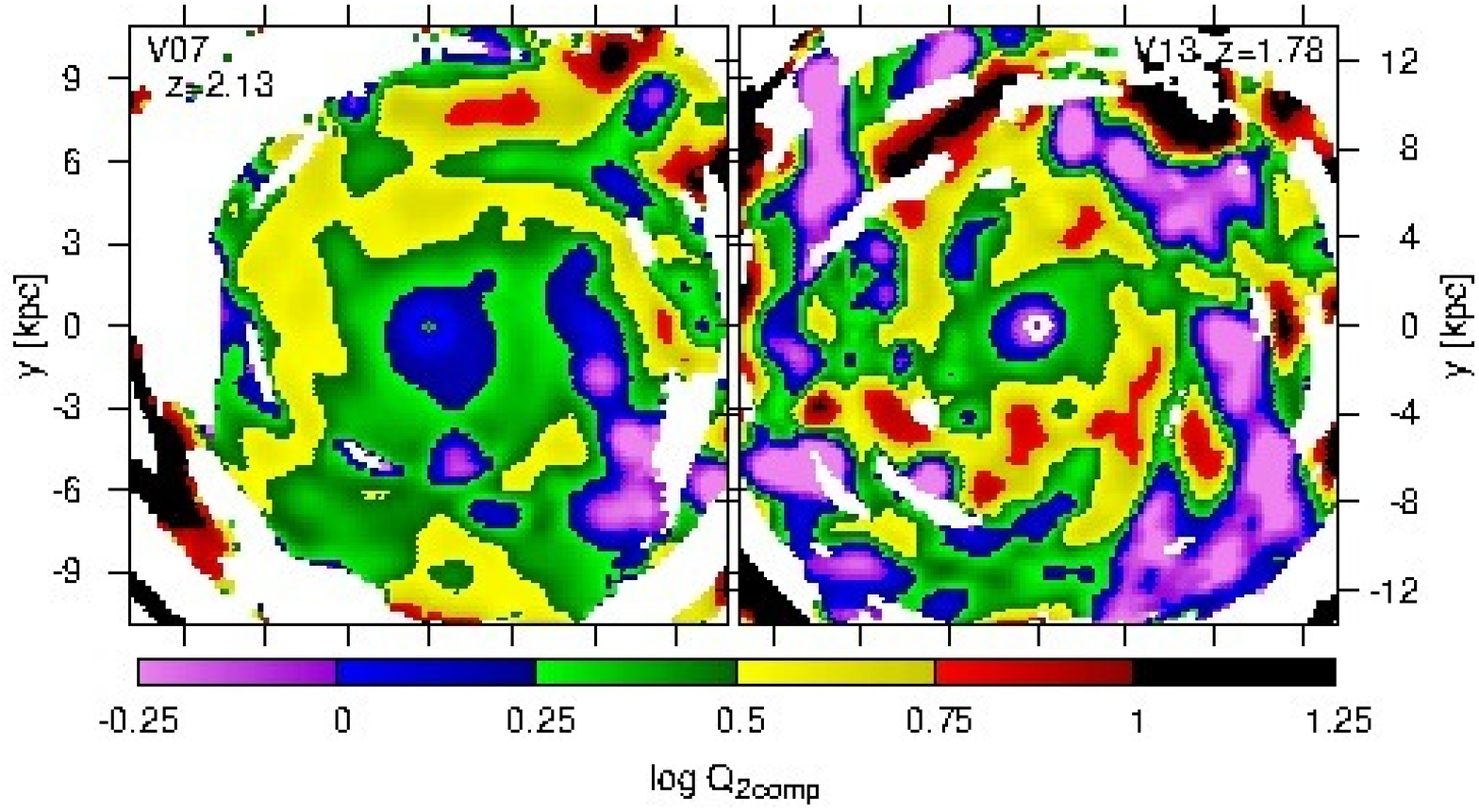}
  \caption{Maps of $Q_{\rm 2comp}$ without applying the bulge-star removal in V07 at $z=2.13$ (left) and V13 at $z=1.78$ (right). The other settings of the analysis and formulation are the same as the results in \S\ref{Results}.}
  \label{WithBulge}
\end{figure}
Fig. \ref{WithBulge} shows the maps of $Q_{\rm 2comp}$ computed without applying the bulge-star removal. Significant decrease of $Q_{\rm 2comp}$ due to the bulge stars can be seen in the central regions of the galaxies, but the values of $Q_{\rm 2comp}$ in the extended disc are hardly changed by the inclusion of the bulge stars. We can say, therefore, that our results of $Q_{\rm 2comp}$ in the disc regions are robust with respect to the removal of the bulge stars.
\begin{figure}
  \includegraphics[width=\hsize]{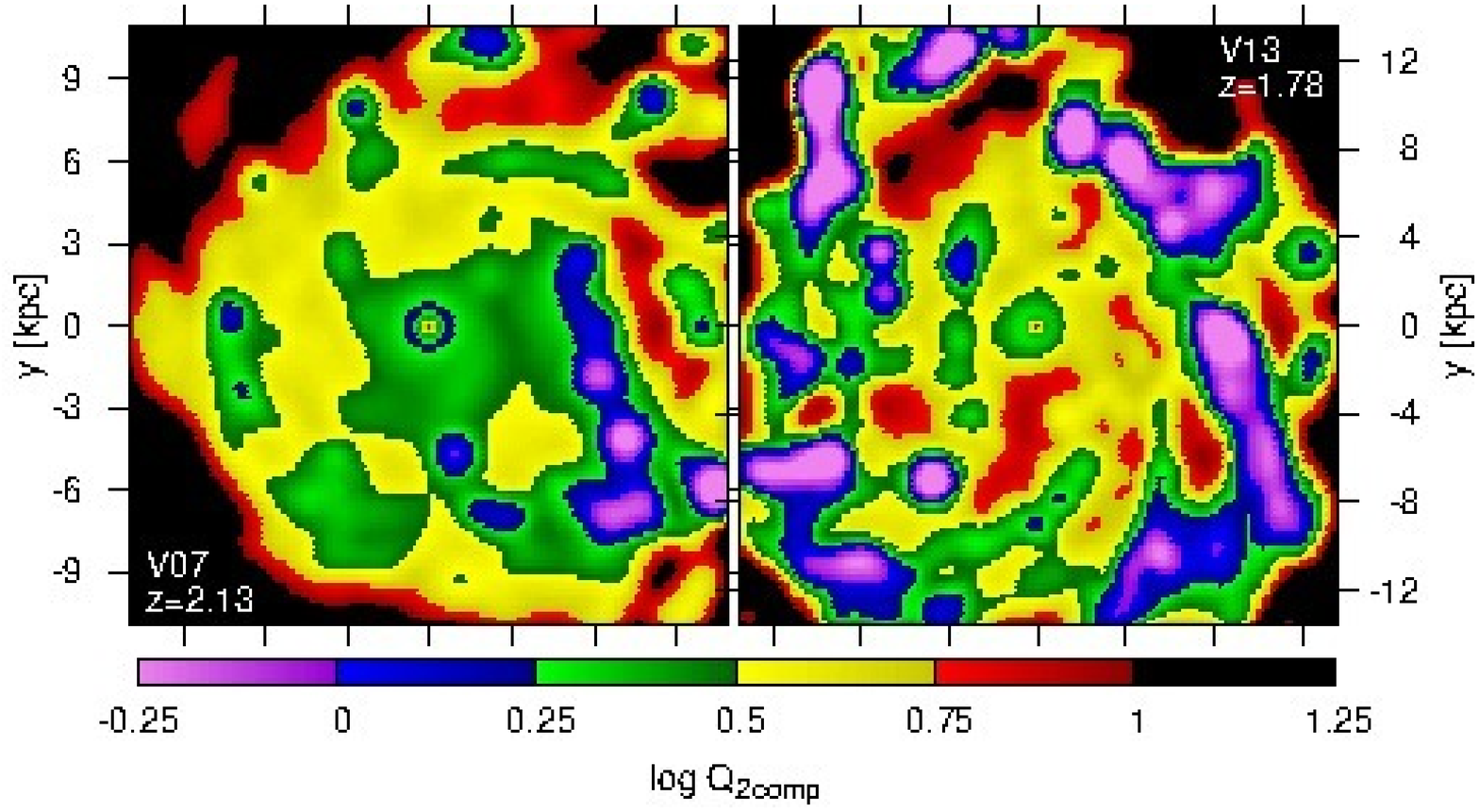}
  \caption{Maps of $Q_{\rm 2comp}$ using circular velocity, $v_{\rm circ}\equiv\sqrt{GM(<r)/r}$, for measuring $\kappa$ and without applying the bulge-star removal in V07 at $z=2.13$ (left) and V13 at $z=1.78$ (right). The other settings of the analysis and formulation are the same as the results in \S\ref{Results}.}
  \label{WithBulge_Vcirc}
\end{figure}
Fig. \ref{WithBulge_Vcirc} shows the maps of $Q_{\rm 2comp}$ obtained without the bulge-star removal and using $v_{\rm circ}$ to measure $\kappa$. The circular velocity $v_{\rm circ}$ is not affected by the exclusion of bulge stars since it is only determined by the gravitational potential. In Fig. \ref{WithBulge_Vcirc}, the values of $Q_{\rm 2comp}$ only slightly increase in the disc regions in comparison with the bottom left panels of Fig. \ref{QMaps_V07_z2.12} and \ref{QMaps_V13_z1.78}. The inter-clump regions generally have $Q_{\rm 2comp}\gsim1.8$. We can say, therefore, that our results of $Q_{\rm 2comp}$ in the disc regions are robust with respect to the removal of the bulge stars.

\subsection[]{Dependence on the cut-off height}
\label{Height}
We test the robustness of our results with respect to the cut-off height. We take into account stars and gas within the height of $|z|<3~{\rm kpc}$ in our analysis. This vertical cut-off is high enough to capture most of the stellar and gas masses belonging to the discs. However, a higher region is expected to have a slower rotation velocity because of velocity lag: usually $\mathrm{d}\overline{v_{\phi}}/\mathrm{d}|z|<0$ \citep{bt:08}. Moreover, $\sigma$ may also vary with height above the plane.

\begin{figure}
  \includegraphics[width=\hsize]{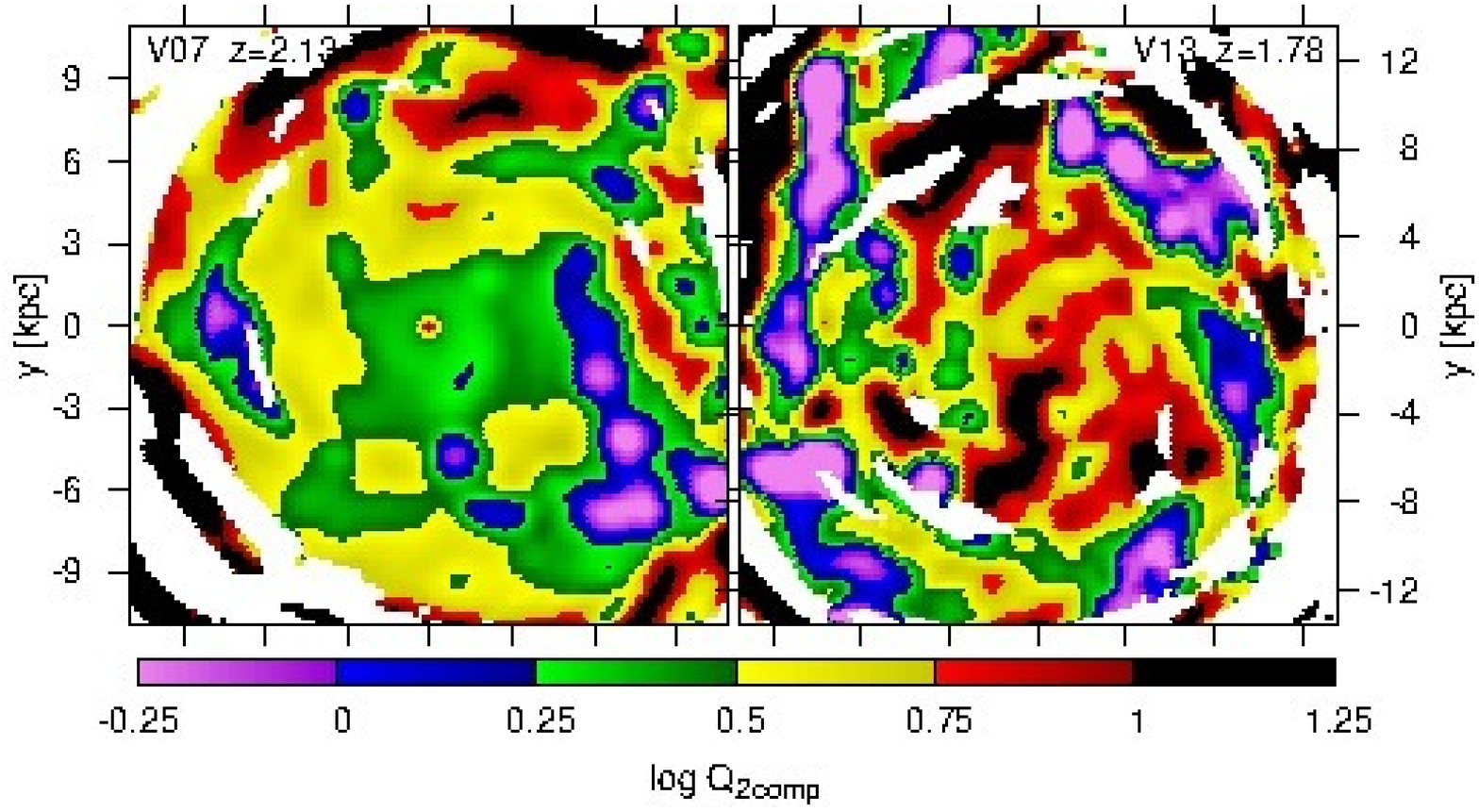}
  \caption{Maps of $Q_{\rm 2comp}$ applying the cut-off height of $|z|<1.5~{\rm kpc}$ in V07 at $z=2.13$ (left) and V13 at $z=1.78$ (right). The other parameters and formulation are the same as the results in \S\ref{Results}.}
  \label{z1.5kpc}
\end{figure}
Fig. \ref{z1.5kpc} shows the maps of $Q_{\rm 2comp}$ with a shorter cut-off height of $|z|<1.5~{\rm kpc}$. Because some clumps are not captured within the thinner slice, some regions indicate higher $Q_{\rm 2comp}$. However, most of regions are hardly affected by the the width of the cut-off height within the range of $|z|=1.5$--$3~{\rm kpc}$.

\subsection[]{Dependence on the smoothing length}
\label{SmoothingDependence}
It should be noted that $Q_{\rm 2comp}$ can depend on physical scales on which the Toomre analysis is applied \citep[][]{rba:10,hr:12,ra:14,arg:15}.\footnote{These previous studies have discussed the dependence of $Q$ values on the Larson-type scaling relations of surface density and velocity dispersion.} In this study, it corresponds to the Gaussian smoothing length that is introduced in \S\ref{smoothing}. A larger smoothing length leads to more uniform distribution of the physical quantities and can cause systematic increase of velocity dispersion if $\sigma$ is not uniform. Using a too large smoothing length, therefore, may overestimate $Q$ values because of the systematic increase of $\sigma$. Furthermore, the smoothing of FWHM$=1.2~{\rm kpc}$ applied in \S\ref{Results} may be too large for the compact discs of V19 and V32 (Fig. \ref{QMaps_V19_z4.88} and \ref{QMaps_V32_z3.00}). 

\begin{figure}
  \includegraphics[width=\hsize]{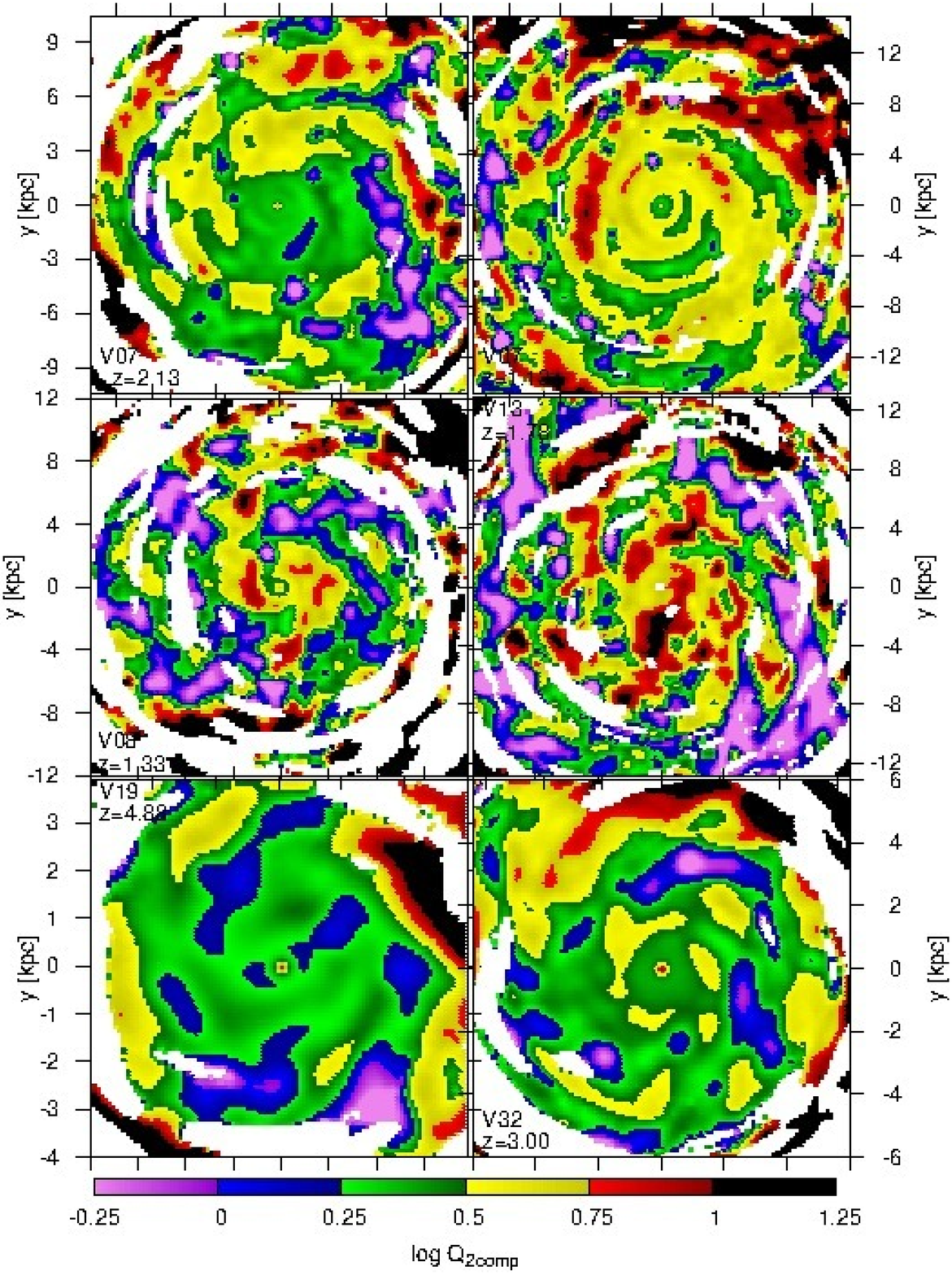}
  \caption{Maps of $Q_{\rm 2comp}$ using the halved Gaussian smoothing length (FWHM$=0.59~{\rm kpc}$). The top panels indicate the results for V07 at $z=2.13$ (left) and $z=1.13$ (right). The middle panels are for V08 (left) and V13 (right). The bottom panels are for V19 (left) and V32 (right).}
  \label{HalfSmoothing}
\end{figure}

\begin{figure}
  \includegraphics[width=\hsize]{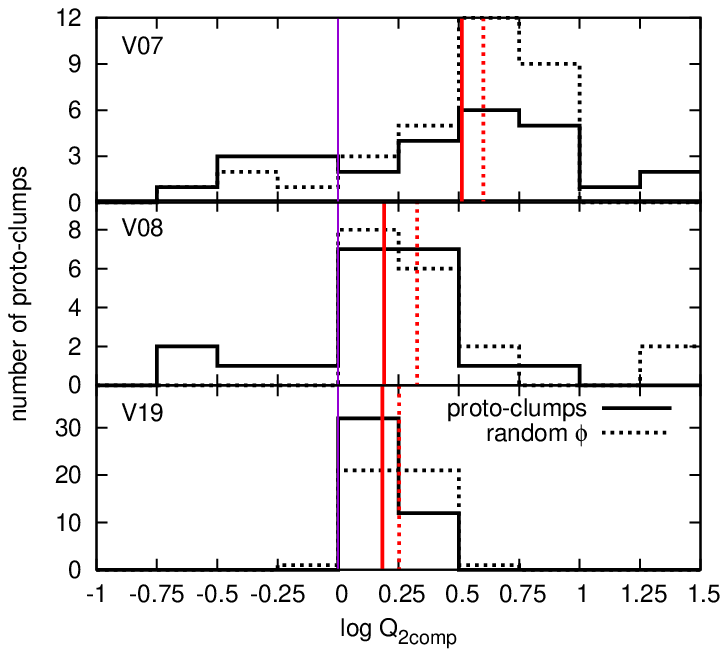}
  \caption{The same as Fig. \ref{Qhist} but with FWHM$=0.59~{\rm kpc}$ for the Gaussian smoothing.}
  \label{Qhist_HS}
\end{figure}

In Fig. \ref{HalfSmoothing}, we show the same maps of $Q_{\rm 2comp}$ but using a Gaussian smoothing kernel with a standard deviation of $0.25~{\rm kpc}$ (FWHM$=0.59~{\rm kpc}$). As seen in the Figure, larger numbers of smaller structures indicating low $Q_{\rm 2comp}$ can be seen than in the cases of the larger smoothing of FWHM$=1.2~{\rm kpc}$. However, the regions showing $Q_{\rm 2comp}$ are still confined to the regions around clumps and do not cover the entire discs. In this sense, our results are robust with respect to the smoothing length in the range between FWHM$=0.59~{\rm kpc}$ and $1.2~{\rm kpc}$, and it can be said that the extended disc regions in our cosmological simulations generally have $Q_{\rm 2comp}\gsim1.8$. Fig. \ref{Qhist_HS} shows the distribution of $Q_{\rm 2comp}$ on the proto-clump and the inter-proto-clump regions (see \S\ref{ThinSlices}) when we apply the shorter Gaussian smoothing of FWHM$=0.59~{\rm kpc}$. The distribution seems to be qualitatively the same as that in the case of FWHM$=1.2~{\rm kpc}$ (Fig. \ref{Qhist}): $Q_{\rm 2comp}\gsim1.8$ on significant fractions of the proto-clump positions. The median values of the distribution (the red lines) are 3.25 (4.00), 1.55 (2.12) and 1.53 (1.79) for the proto-clump (inter-proto-clump) regions in V07, V08 and V19, respectively. We might find a larger number of unstable regions with $Q<1$ if we use even smaller smoothing, and such small-scale instability would result in formation of less massive clumps of $\lsim10^7~{\rm M_\odot}$. Although the small clumps may grow to massive ones by merging with other clumps and/or rapid accretion of surrounding gas, this effect is beyond the applicable domain of the linear perturbation theory (see \S\ref{Possibility2}). 

\end{document}